\documentclass[twocolumn,aps,prx,showpacs,amsmath,superscriptaddress,longbibliography,notitlepage]{revtex4-2}
\usepackage{amssymb}
\usepackage{mathrsfs}
\usepackage{graphicx}
\usepackage{float}
\usepackage[caption=false]{subfig}
\usepackage[pdftex,colorlinks,linkcolor={red!75!black},citecolor={red!75!black},urlcolor={blue!75!black}]{hyperref}
\usepackage{verbatim}
\usepackage{epstopdf}
\usepackage[normalem]{ulem}
\usepackage{verbatim}
\usepackage{xcolor}
\usepackage{bm}

\newcommand{\clr}{\color{red!75!black}}

\newcommand{\Rnum}[1]{\uppercase\expandafter{\romannumeral #1\relax}}

\begin{document}

\title{Point-Gap Bound States in Non-Hermitian Systems} 
\author{Zixi Fang}
\affiliation{Beijing National Laboratory for Condensed Matter Physics, and Institute of Physics, Chinese Academy of Sciences, Beijing 100190, China}
\affiliation{University of Chinese Academy of Sciences, Beijing 100049, China}
\author{Chen Fang}
\email{cfang@iphy.ac.cn}
\affiliation{Beijing National Laboratory for Condensed Matter Physics, and Institute of Physics, Chinese Academy of Sciences, Beijing 100190, China}
\affiliation{Songshan Lake Materials Laboratory, Dongguan, Guangdong 523808, China}
\affiliation{Kavli Institute for Theoretical Sciences, Chinese Academy of Sciences, Beijing 100190, China}
\author{Kai Zhang}
\email{phykai@umich.edu}
\affiliation{Beijing National Laboratory for Condensed Matter Physics, and Institute of Physics, Chinese Academy of Sciences, Beijing 100190, China}
\affiliation{Department of Physics, University of Michigan Ann Arbor, Ann Arbor, Michigan, 48109, United States}

\begin{abstract}
In this paper, we systematically investigate the impurity-induced bound states in 1D non-Hermitian systems. 
By establishing an exact relationship between impurity potential and bound-state energy, we determine the minimum impurity potential required to generate bound states within each point energy gap. 
We demonstrate that the absence of Bloch saddle points necessitates a finite threshold of impurity potential; otherwise, infinitesimal impurity potential can create bound states. 
Furthermore, we show that the bound states residing in the point gaps with nonzero spectral winding exhibit sensitivity to boundary conditions and will be squeezed towards the edges when the boundaries are opened, indicating the bulk-boundary correspondence in terms of point-gap topology. 
\end{abstract}

\maketitle

\emph{\clr Introduction.---}~When a system interacts with external environments, such as optical systems with balanced gain and loss~\cite{Regensburger2012,Gao2015_Nature,FengLiang2017,Ganainy2018,Miri2019,YangLan2019} or quasi-particle excitations with a finite lifetime~\cite{FuLiang2017_arXiv,ShenHT2018_PRL,FuLiang2020_PRL}, its non-unitary time evolution or broadened spectral function necessitates an effective non-Hermitian Hamiltonian description~\cite{Torres2018Review,Kunst2019,Ashida2020,DingKun2022_NRP,GongPRX2018,Sato2019_PRX}. 
Recently, many intriguing phenomena have appeared in non-Hermitian band systems, with a major focus on the non-Hermitian skin effect (NHSE)~\cite{Lee2016PRL,luisPRB2018,Yao2018,Kunst2018_PRL,WangZhong2018,Murakami2019_PRL,ChingHua2019,LeeCH2019_PRL,LonghiPRR2019,Kai2020,Okuma2020_PRL,Zhesen2020_aGBZ,Zhesen2020_SE,XuePeng2020,Ghatak2020,Thomale2020,LiLH2020_NC,Kawabata2020_Symplectic,Wanjura2020_NC,XueWT2021_PRB,LiLH2021_NC,Kai2022NC,ZhangDDS2022,Longhi2022PRL}. 
This effect is characterized by the condensation of the majority of system eigenstates on its boundary when open boundary conditions (OBCs) are imposed.
These localized eigenstates can be quantitatively described using the generalized Brillouin zone (GBZ) that consists of complex momenta~\cite{Yao2018,Murakami2019_PRL,Kai2020,Zhesen2020_aGBZ}. 

Impurities are common in realistic materials and significantly responsible for their transport properties~\cite{PeresRMP}. 
For instance, doping impurities in semiconductors enhance conductivity~\cite{PantelidesRMP}, and magnetic impurities in metals lead to Kondo effect~\cite{Kondo2964}. 
This importance motivates the investigation of impurity states in non-Hermitian systems~\cite{Hatano1996,Hatano1997_PRB,li2021impurity,liu2020diagnosis,LiuYX2020_PRB,LiuYX2021_PRB,roccati2021non,gong2022anomalous}. 
It was known that NHSE manifests as a boundary phenomenon in non-Hermitian systems, and impurities with codimension-1 can be treated as a soft boundary~\cite{Hsu2016Review,ZhangDDS2022}. 
Given this connection, it is natural to ask whether NHSE influences the formation of impurity bound states. 
Moreover, impurities serving as soft boundaries can be applied to investigate the bulk-boundary correspondence~\cite{Kane2010,ChiuCK2016_RMP,Slager2015_PRB,ChiuCK2016_RMP,SauJD2013,Meier2016,Diop2020PRB,BorgniaDS2020_PRL}.
In non-Hermitian band systems, the complex eigenvalues enrich the types of energy gaps~\cite{GongPRX2018,Sato2019_PRX}. 
It allows us to assign a spectral winding number to each point gap when the periodic boundary condition (PBC) is applied. 
Previous studies revealed that NHSE has a point-gap topological origin~\cite{Kai2020,Okuma2020_PRL}, i.e., the collapse of a point gap with a nonzero spectral winding number leads to the emergence of NHSE under OBCs. 
Therefore, an additional question arises whether bound states in point gaps can reflect the bulk-boundary correspondence regarding point-gap topology in non-Hermitian systems. 

In this paper, we present a general theory of impurity bound states in 1D non-Hermitian lattice systems, utilizing Green's function method to establish the exact relation between the strength of impurity potential and the corresponding bound-state energy.  
We reveal that in the absence of Bloch saddle points~\cite{ashcroft2022solid}, a finite impurity potential threshold is required to generate bound states; otherwise, an infinitesimal impurity potential can produce bound states, indicating the crucial role of Bloch saddle points in determining the minimum impurity potential for producing bound states. 
Here, the Bloch saddle points refer to the momenta $k_s$ satisfying $\partial_{k}\mathcal{H}(k_s)=0$. 
Meanwhile, when NHSE is present, the envelope of the bound state exhibits asymmetric localization away from the impurity site, as shown in Fig.~\ref{fig:1}(d). 
Furthermore, we demonstrate that a single impurity can confine multiple bound states, among which the ones residing in the point gaps with nonzero spectral winding are sensitive to the boundary conditions and will be pushed onto the edges when the OBCs are imposed, suggesting the presence of NHSE. 

\emph{\clr The formation of bound states in non-Hermitian systems}.---
We start with a general single-band tight-binding model under PBCs, of which the Hamiltonian can be expressed as,
\begin{equation}\label{MT_GenFreeHam}
\begin{split}
	H_0 = \sum_{r}\sum_{l=-m}^{n} t_{l} |r\rangle \langle r+l|  = 
	\sum_{k\in \mathrm{BZ}} \mathcal{H}_0(z\equiv e^{ik}) |k\rangle\langle k|
\end{split}
\end{equation}
with $\mathcal{H}_0(z) =  \sum_{l=-m}^{n} t_l \, z^l $ a Laurent polynomial of $z$, $r$ representing the lattice site, and $t_l$ indicates the hopping parameter that only depends on the (finite) hopping range $l$ due to the translation symmetry. 
As momentum $k$ transverses the entire BZ, the corresponding PBC spectrum forms arcs or loops, denoted as $\sigma_{\mathrm{PBC}}$, which divides the complex energy plane into several disconnected regions called point gaps~\cite{Sato2019_PRX,DingKun2022_NRP} and labeled as $\mathcal{E}_0, \mathcal{E}_1, \dots$. 
We always specify $\mathcal{E}_0$ as the point gap connected to infinity, as illustrated in Fig.~\ref{fig:1}(a)(c).

\begin{figure}[t]
	\begin{center}
		\includegraphics[width=1\linewidth]{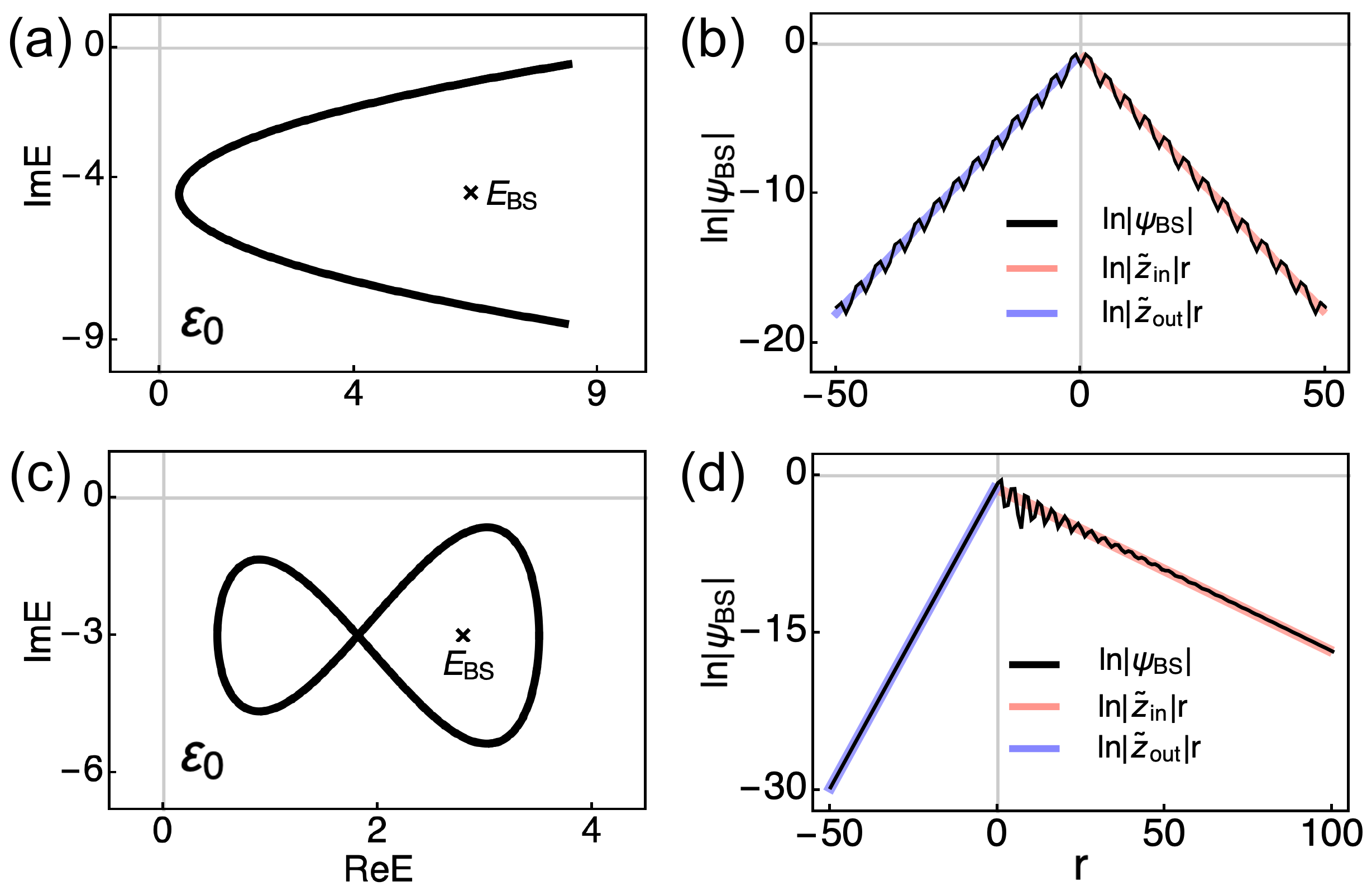}
		\par\end{center}
		\protect\caption{\label{fig:1} 
		The parameters of the Hamiltonian in Eq.(\ref{MT_ModelHam}) $\{t_{-2},t_{-1},t_{1},t_{2},u\}$ are chosen as $\{-2,-2,2,-2,4.4-4.5 i\}$ for (a)(b) and $\{-1,1/2,1,1,2-3i\}$ for (c)(d). 
		(a)(c) represent the PBC spectra (the black curves) as well as the bound state energy $E_{\mathrm{BS}}$ located in the point gaps (the black asterisk). 
		(b)(d) show the logarithm of two bound states (the black lines) and the comparison with two straight lines, $\ln|\tilde{z}_{\mathrm{in}}|r$ in the $r>0$ region (the red line) and $\ln|\tilde{z}_{\mathrm{out}}|r$ in the $r<0$ region (the blue line). 
		} 
\end{figure}

Consider a single impurity at the center of the 1D periodic chain, 
\begin{equation}\label{MT_Potential}
	V =\lambda \, \sum\nolimits_r \delta(r)  |r\rangle \langle r|,
\end{equation}
where $\lambda$ represents the strength of the impurity potential and assumes a complex value~\cite{Ganainy2018}. 
The single impurity potential can produce bound states that are localized around the impurity site and have energies within point gaps, as shown in Fig.~\ref{fig:1}(a)(c). 
The eigenfunction $\psi_E(r)$ with energy $E$ in the full Hamiltonian $H=H_0+V$ can be determined using Green's function~\cite{SupMat}, that is, $\psi_E(r) = \lambda \, \psi_E(0) \, G_0(E; r)$, where $G_0(E;r)=\langle r | 1/(E-H_0)|0\rangle$ with $H_0$ under PBC. 
Here, $\psi_E(0)$ is known by the normalization condition of $\psi_E(r)$. 
Based on the Hamiltonian in Eq.(\ref{MT_GenFreeHam}), the Green's function can be further expressed in an integral form, 
\begin{equation}\label{MT_GreenFunc}
	G_0(E; r) =  \oint_{C} \frac{dz}{2 \pi i} \frac{z^{r+m-1}}{P_{E}(z)},
\end{equation}
where $M$ indicates the multiplicity of the pole in $\mathcal{H}_0(z)$ and $P_{E}(z)=z^{m}(E - \mathcal{H}_0(z))$ represents an non-negative-order polynomial with respect to $z$ for a given energy $E$. 
Under PBC, the integral contour $C$, namely BZ, is the unit circle $|z|=1$ in the complex $z$ plane. 
Therefore, for a given point-gap energy $E_{\mathrm{BS}}$ produced by the impurity potential of strength $\lambda$, the corresponding bound state can be calculated as 
\begin{equation}\label{MT_BoundStates}
	\psi_{\mathrm{BS}}(r) = \lambda \, \psi_{\mathrm{BS}}(0)
	\left\{  \begin{split}
	&\sum_{|z|<1} {\mathrm{R}}(E_{\mathrm{BS}}, z) \, z^{r}, \,\,\,\,\,\,\,\,\, r > 0; \\  
	&\sum_{|z|>1} - {\mathrm{R}}(E_{\mathrm{BS}}, z) \,  z^{r}, \,\,\,\, r < 0,
	\end{split}  \right.  
\end{equation}
where $\mathrm{R}(E_{\mathrm{BS}}, z_i)=-z_i^{m-1}/t_{n} \, \Pi_{j(\neq i) =1}^{m+n} 1/(z_i-z_j)$ is the residue of the function $[z(E_{\mathrm{BS}} - \mathcal{H}_0(z))]^{-1}$ at its pole $z_i$. 
Note that these poles correspond to zeros of $P_{E_{\mathrm{BS}}}(z)$ in Eq.(\ref{MT_GreenFunc}). 
The formula of bound states in multi-band cases are discussed in~\cite{SupMat}. 
It can be derived from Eq.(\ref{MT_BoundStates}) that bound states exhibit exponential localization; when away from the impurity site, the localization behavior on the right (left) side of the impurity is dominated by the largest (smallest) poles inside (outside) $|z|=1$. 

Here, we consider two examples, one without NHSE (Fig.~\ref{fig:1}(a)(b)) and another with NHSE (Fig.~\ref{fig:1}(c)(d)), to show that the presence of NHSE results in asymmetric decay behaviors for the bound states. 
The model Hamiltonian is composed of the free part
\begin{equation}\label{MT_ModelHam}
	\mathcal{H}_0(z) = t_{-2}z^{-2}+t_{-1}z^{-1} + t z + t_2 z^2 + u 
\end{equation}
and impurity potential in Eq.(\ref{MT_Potential}) with strength $\lambda$. 
The bound states are created with impurity potential $\lambda=15$ in Fig.~\ref{fig:1}(a) and $\lambda=5.5$ in Fig.~\ref{fig:1}(c). 
To better characterize the localization behavior, we plot $\ln|\psi_{\mathrm{BS}}(r)|$ for the bound state in each case, marked by the black lines in Fig.~\ref{fig:1}(b) and (d), which matches well with the straight lines of slopes $\ln |\tilde{z}_{\mathrm{in}}|$ in the $r>0$ region (the red line) and $\ln|\tilde{z}_{\mathrm{out}}|$ in the $r<0$ region (the blue line). 
Here, $|\tilde{z}_{\mathrm{in}}|$ ($|\tilde{z}_{\mathrm{out}}|$) represents the largest (smallest) magnitude of the poles within (out) $|z|=1$. 
The slope of the bound state, $\partial_r \mathrm{\ln}|\psi_{\mathrm{BS}}(r)|$, is nothing but the inverse of decay length and dominated by these two dominant poles. 
As a result, the comparison between Fig.~\ref{fig:1}(b) and (d) demonstrates that in the case of NHSE, the bound state exhibits an asymmetric decay behavior away from the impurity. 

\emph{\clr Nonzero threshold of impurity potential for bound states}.---
To further investigate the bound states in different point gaps, we establish the exact relationship between single-impurity potential $\lambda$ and its bound-state energy $E_{\mathrm{BS}}$. 
Remarkably, we reveals that the absence of the Bloch saddle point (BSP) necessitates a finite threshold of impurity potential for the formation of bound states; otherwise, an infinitesimal impurity potential can excite bound states. 

\begin{figure}[t]
	\begin{center}
		\includegraphics[width=1\linewidth]{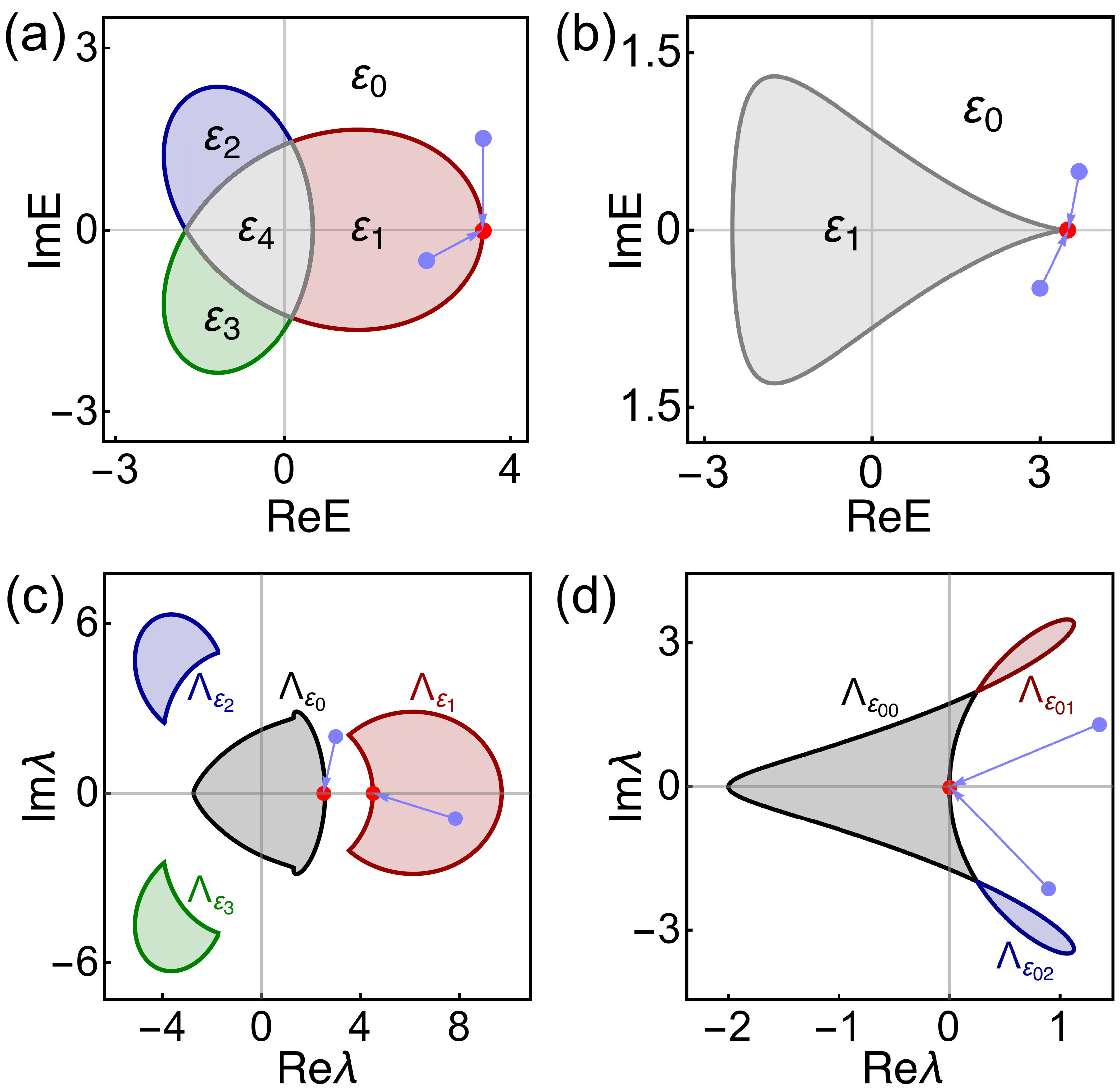}
		\par\end{center}
	\protect\caption{\label{fig:2} 
		Comparison of two examples of (a)(c) without BPS and (b)(d) with BPS.
		The parameters $\{t_{-2},t_{-1},t_{1}\}$ of the Hamiltonian in Eq.(\ref{MT_ModelHam}) are chosen as $\{2,1/2,1\} $ in (a)(c) and $\{1/2,1,2\} $ in (b)(d), respectively. Other parameters are set to be zero. (a)(b) show the PBC spectra and disjoint point gaps indicated by different colors, and (c)(d) present the corresponding $\lambda$ diagrams for bound states.}
\end{figure}

One can obtain the relation between impurity potential $\lambda$ and the corresponding bound-state energy $E_{\mathrm{BS}}$ from Eq.(\ref{MT_BoundStates})~\cite{SupMat}, that is, 
\begin{equation}\label{MT_LambdaSpec}
	\lambda^{-1}(E_\mathrm{BS}) = \sum\nolimits_{|z_i|<1} \mathrm{R}(E_\mathrm{BS}, z_i),
\end{equation}
which gives the strength of impurity potential $\lambda$ required for producing the bound state with energy $E_{\mathrm{BS}}$. 
Since the bound-state energy $E_{\mathrm{BS}}$ is not included in the $\sigma_{\mathrm{PBC}}$, there are no poles touching $|z|=1$, ensuring the above relation always well-defined. 
For each point gap $\mathcal{E}_i$, we can assign a spectral winding number~\cite{Kai2020} for $\mathcal{H}_0(z)$ regarding the bound-state energy, 
\begin{equation}\label{MT_SpecWinding}
	\forall E_{\mathrm{BS}}\in \mathcal{E}_i; \, 
	w_{\mathrm{BZ},\mathcal{E}_i} = n_{z} - n_{p},
\end{equation}
where $n_p=m$ is the multiplicity of the pole in $\mathcal{H}_0(z)$, and $n_z$ is the number of zeros of $E_{\mathrm{BS}}-\mathcal{H}_0(z)$ inside the BZ, depending on the choice of bound-state energy $E_{\mathrm{BS}}$. 
Note that the zeros exactly correspond to the counted poles in Eq.(\ref{MT_LambdaSpec}). 
Therefore, when $E_{\mathrm{BS}}$ lies in the point gaps $\mathcal{E}_i$ with $w_{\mathrm{BZ},\mathcal{E}_i}=n$ or $-m$, all poles are included either inside or outside the trajectory $|z|=1$, which causes the right-hand side of Eq.(\ref{MT_LambdaSpec}) to vanish and requires an infinite impurity potential to create bound states. 
Here, $n$ and $m$ represent the longest hopping range to the left and right in the Hamiltonian $\mathcal{H}_0(z)$, respectively. 
Therefore, we reach the first conclusion: the bound states cannot be created within point gaps that possess spectral winding of $n$ and $-m$. 
Two examples are presented in Fig.~\ref{fig:2}(a)(c) and (b)(d), respectively. 
The point gaps where bound states cannot appear are labeled as $\mathcal{E}_4$ ($w_{\mathrm{BZ},\mathcal{E}_4} = -m= -2$) in Fig.~\ref{fig:2}(a) and $\mathcal{E}_1$ ($w_{\mathrm{BZ},\mathcal{E}_1} = n= 1$) in Fig.~\ref{fig:2}(b). 

We now examine the minimum impurity potentials that can yield bound states within different point gaps. 
The minimum bound-state energy shall be in the point gaps and close to PBC spectrum $\sigma_{\mathrm{PBC}}$. 
We begin with $E_{\mathrm{BS}} \in \mathcal{E}_i$ and let it approach $E \in \sigma_{\mathrm{PBC}}$, as illustrated in Fig.~\ref{fig:2}(a)(b), the corresponding impurity potential $\lambda(E_{\mathrm{BS}})$ in Eq.(\ref{MT_LambdaSpec}) reaches a limit value, as shown in Fig.~\ref{fig:2}(c)(d). 
Likewise, for each point gap $\mathcal{E}_i$, we can define the set of minimum impurity potentials required to create bound states in this point gap, 
\begin{equation}\label{MT_Limitation}
	\Lambda_{\mathcal{E}_i}:=\{ \lim\nolimits_{E_{\mathrm{BS}}\rightarrow E} \lambda(E_{\mathrm{BS}}) | E_{\mathrm{BS}} \in \mathcal{E}_i,  E \in \sigma_{\mathrm{PBC}} \}.
\end{equation}
As illustrated in Fig.~\ref{fig:2}(a), there are four disjoint point gaps $\mathcal{E}_{i=0,1,2,3}$ that allow for bound states. 
Relating to these point gaps, we identify four sets of minimum impurity potentials derived from Eq.(\ref{MT_Limitation}), $\Lambda_{\mathcal{E}_{i=0,1,2,3}}$, i.e., four different colored boundary curves in Fig.~\ref{fig:2}(c). 
When the impurity potential $\lambda$ is inside the gray region surrounded by $\Lambda_{\mathcal{E}_{0}}$ in Fig.~\ref{fig:2}(c), it is insufficiently strong to give rise to bound states. 
As $\lambda$ surpasses the boundary $\Lambda_{\mathcal{E}_{0}}$, bound states first appear in the point gap $\mathcal{E}_0$. 
As shown in Fig.~\ref{fig:2}(c), generating bound states in other point gaps necessitates even larger impurity potentials. 
If $\lambda$ goes into the colored region, e.g., the red region encircled by $\Lambda_{\mathcal{E}_{1}}$, multiple bound states are produced, one in the point gap $\mathcal{E}_{1}$ and another (not shown in Fig.~\ref{fig:2}(a)) in the $\mathcal{E}_{0}$. 
More details about the $\lambda$ diagram are presented in~\cite{SupMat}. 
Therefore, the finite gray area covering the origin $\lambda=0$ in Fig.~\ref{fig:2}(c) indicates a finite threshold of impurity potentials required to generate point-gap bound states. 

While Hamiltonian $H_0(z)$ exhibits BSP, an infinitesimal impurity potential can excite bound states. 
We define the BSP as the saddle point with a unit modulus that satisfies $\partial_z \mathcal{H}_0(z)|_{z=z_s}=0$ and $|z_s|=1$ simultaneously.
The energy at BSP is denoted as $E_s=\mathcal{H}_0(z_s)$. 
One example with BSP is shown in Fig.~\ref{fig:2}(b)(d). 
When $E_{\mathrm{BS}}$ approaches $E_s\in \sigma_{\mathrm{PBC}}$, as illustrated in Fig.~\ref{fig:2}(b), there are at least two poles in Eq.(\ref{MT_LambdaSpec}), denoted as $z_{s1}$ inside $|z|=1$ and $z_{s2}$ outside $|z|=1$, that are closest to each other with the distance $\delta z_s=|z_{s1}-z_{s2}|$ and coincide exactly at BSP ($\delta z_s=0$)~\cite{SupMat}. 
Consequently, the inverse of impurity potential $\lambda^{-1}(E_{\mathrm{BS}})$ in Eq.(\ref{MT_LambdaSpec}) is dominated by the residue $R(E_{\mathrm{BS}},z_{s1})\propto 1/\delta z_s$. 
Based on Eq.(\ref{MT_Limitation}), the minimum impurity potential is attained at BSP energy, $\Lambda_{\mathcal{E}_0}(E_s) = \lim_{E_{\mathrm{BS}} \rightarrow E_s} \lambda(E_{\mathrm{BS}}) \approx \delta z_s \rightarrow 0$, which means that an infinitesimal impurity potential $\lambda$ can excite the bound states with energies near $E_s$. 
The set of minimum impurity potentials $\Lambda_{\mathcal{E}_0}$ for the point gap $\mathcal{E}_0$ is shown in Fig.~\ref{fig:2}(d), which is composed of three pieces $\Lambda_{\mathcal{E}_{00}}, \Lambda_{\mathcal{E}_{01}}$, and $\Lambda_{\mathcal{E}_{02}}$ connected by two self-intersections. 
Thus, zero threshold of impurity potential means the curve $\Lambda_{\mathcal{E}_0}$ crosses $\lambda=0$, as shown in Fig.~\ref{fig:2}(d). 
More details for this example are presented in~\cite{SupMat}.
Comparing with the case in Fig.~\ref{fig:1}(a)(c), we conclude: when the Hamiltonian lacks BSP, a finite impurity potential is needed to excite bound states; otherwise, an infinitesimal impurity potential can produce bound states with energies near BSP energy. 

\emph{\clr The sensitivity of point-gap bound states to boundary conditions}.---
Here, we demonstrate that these point-gap bound states exhibit sensitivity to boundary conditions and are driven towards the boundaries following the collapse of point gap at that bound-state energy. 

The transition between boundary conditions can be parameterized by the boundary link strength $s$ under the Hamiltonian $H_{0}^{s}=H_0- s H_{B}$, where $H_0$ indicates the PBC Hamiltonian and $H_B$ represents the boundary hopping terms. 
As the parameter $s$ goes from $0$ to $1$, the Hamiltonian $H_0^s$ is modulated from PBC to OBC; correspondingly, the integral contour $C$ in Eq.(\ref{MT_GreenFunc}) undergoes continuous deformation from BZ into GBZ with the intermediary trajectory denoted as $C_{s}$~\cite{SupMat}. 
In this process, the trajectory $C_s$ sweeps through the blue shaded region shown in Fig.~\ref{fig:3}(b), that is the difference between the interiors of BZ and GBZ. 
We label the intermediary spectrum as $\sigma_s:=\{ \mathcal{H}_0(z)|z\in C_s, 0<s<1 \}$ and OBC spectrum as $\sigma_{\mathrm{OBC}}$. 
After introducing the impurity potential $V$ defined in Eq.(\ref{MT_Potential}), the total Hamiltonian becomes $H^s = H_0^s+V$, and impurity states can appear.  
Here we define the inverse of decay length of impurity state with energy $E$ on the right side of impurity site as 
\begin{equation}\label{MT_kappa}
	\begin{aligned}
		& \kappa_{+}:=\partial_r \ln \left|\psi_E(r)\right|; \quad r>0.
	\end{aligned}
\end{equation}
Under the boundary condition with boundary link $s$, it can be derived from Eq.(\ref{MT_BoundStates}) that the localization behavior of the impurity state with energy $E$ in is dominated by the largest poles included by $C_s$ instead of $|z|=1$.  
Meanwhile, the relation between bound-state energy and requisite impurity potential can be obtained from Eq.(\ref{MT_LambdaSpec}), where the residue is summed over the poles within the interior of trajectory $C_s$ rather than BZ. 

An illustrative example is shown in Fig.~\ref{fig:3}, and the Hamiltonian under PBC ($s=0$) is the same as that in Fig.~\ref{fig:2}(a). 
We start from two bound states within different point gaps. 
One bound state with energy marked as $E_1$ (the red cross) in Fig.~\ref{fig:3}(a) resides in the point gap $\mathcal{E}_1$. 
Two poles of $[E_1-\mathcal{H}_0(z)]^{-1}$ within GBZ are labeled as $z_{1,2}$ (the red dots in Fig.~\ref{fig:3}(b)). 
The spectral winding number $w_{\mathrm{BZ,\mathcal{E}_1}}=-1$ in Eq.(\ref{MT_SpecWinding}) means that only one pole $z_2$ is located in the intermediary region between BZ and GBZ. 
As the boundary conditions change from PBC to OBC, $C_s$ deviates from BZ, traverses the pole $z_2$ within the shaded region ($s=s_c$) shown in Fig.~\ref{fig:3}(b), and eventually enters GBZ. 
Before and after the transition ($s=s_c$), the dominant pole enclosed by $C_s$ will be changed. 
As a result, $\kappa_+$ for the bound state with energy $E_1$ experiences an jump, as shown in Fig.~\ref{fig:3}(c). 
Correspondingly, the bound state $\psi_{\mathrm{BS}}(r)$ transitions into an edge mode with a diverging amplitude ($\kappa_+>0$) under large-size limit, as illustrated by the red profile in the insets of Fig.~\ref{fig:3}(c). 
It exhibits that the bound state in the point gap with nonzero spectral winding is unstable and sensitive to the boundary conditions. 
In contrast, the bound state with energy $E_0$, labeled by the blue cross in Fig.~\ref{fig:3}(a), stays in the point gap $\mathcal{E}_0$. 
Due to the zero spectral winding number of $\mathcal{E}_0$, the trajectory $C_s$ does not across any (blue) poles for $0\leq s \leq 1$, as shown in Fig.~\ref{fig:3}(b). 
Therefore, this point-gap bound state is stable to the change in boundary conditions, indicated by the blue profile in the insets of Fig.~\ref{fig:3}(c). 

\begin{figure}[t]
	\begin{center}
		\includegraphics[width=1\linewidth]{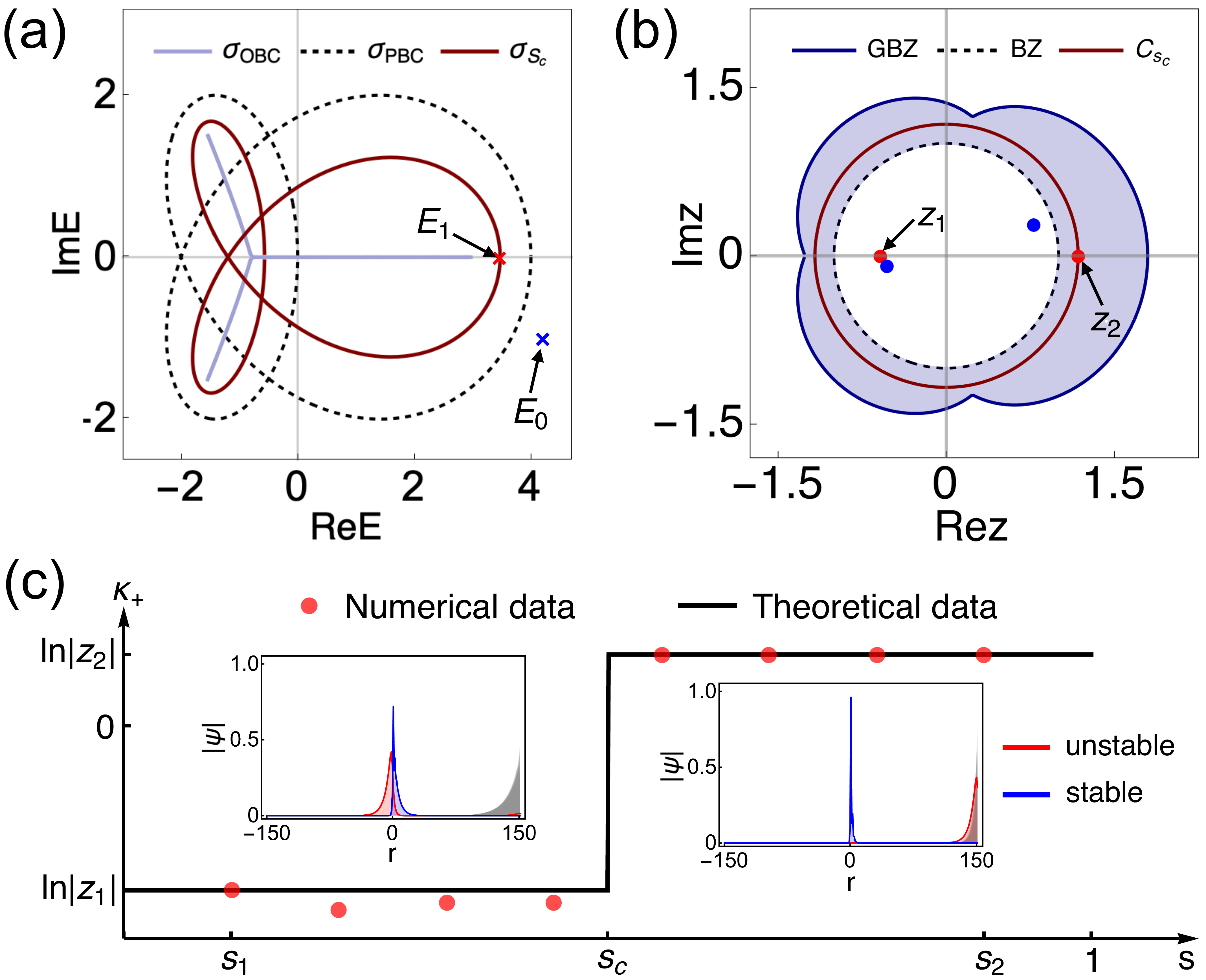}
		\par\end{center}
	\protect\caption{\label{fig:3} 
		(a) The spectra under different boundary conditions and the two bound-state energies $E_0$ and $E_1$, marked by the blue and red crosses. 
		(b) BZ, GBZ, and critical trajectory $C_{s_c}$ that crosses the pole $z_2$. The red and blue dots denote the poles related to $E_1$ and $E_0$, respectively. 
		(c) The inverse of decay length $\kappa_+$ for $\psi_{E_1}(r)$ experiences one jump at $s=s_c$. 
		Two insets show the spatial profiles of impurity states with the boundary link $s_1=1-10^{-9}$ and $s_2=1-10^{-30}$, respectively.
		For comparison, the bulk states shown by the gray profiles are localized on the edge once $s\neq 0$. 
		The system parameters in $\mathcal{H}_0(z)$ are the same as Fig.~\ref{fig:2}(a). }
\end{figure}

In general cases, we start from a bound state $\psi_{\mathrm{BS}}(r)$ with the energy $E_{\mathrm{BS}}$ inside the point gap $\mathcal{E}_i$ under PBC ($s=0$), and examine its localization behavior as the boundary conditions vary from PBC to OBC. 
Suppose that the point gap has a spectral winding $w_{\mathrm{BZ,\mathcal{E}_i}}=q$, as defined in Eq.(\ref{MT_SpecWinding}), the BZ contains $n_p+q$ poles in Eq.(\ref{MT_LambdaSpec}), while GBZ always encloses $n_p$ poles~\cite{Kai2020}. 
There are $q$ poles distributed in the intermediary region between BZ and GBZ. 
As the boundary condition varies from PBC ($s=0$) to OBC ($s=1$), the spectrum $\sigma_{s}$ traverses the energy $E_{\mathrm{BS}}$ $q$ times before finally collapses into OBC spectrum $\sigma_{\mathrm{OBC}}$. 
Accordingly, the trajectory $C_s$ sweeps $q$ poles as it deforms from BZ into GBZ, resulting in the inverse of decay length $\kappa_+$ experiencing $q$ abrupt changes. 
One example with $q=2$ is shown in~\cite{SupMat}. 
The number of jumps in $\kappa_{+}$ exactly corresponds to the spectral winding number $w_{\mathrm{BZ,\mathcal{E}_i}}=q$ of the point gap, indicating the bulk-edge correspondence regarding the point-gap topology. 

\emph{\clr Conclusions.---}~In summary, we investigate the interplay between point gaps and bound states induced by a single impurity within 1D non-Hermitian lattice systems. 
We establish the exact relationship between bound state energy and required impurity potential and reveal the critical role of BSP on the minimum threshold of impurity potentials. 
Specifically, in the absence of BSP, a finite impurity potential is required for the generation of bound states; otherwise, an infinitesimal potential is sufficient to create bound states with energies close to that of the BSP. 
Meanwhile, we show that the presence of NHSE causes the asymmetric localization length of bound states. 
We demonstrate that the bound states within the point gaps characterized by nonzero spectral winding are unstable and sensitive to boundary conditions. 
The sensitivity of point-gap bound states indicates the bulk-edge correspondence of point-gap topology specific to non-Hermitian systems. 

\let\oldaddcontentsline\addcontentsline
\renewcommand{\addcontentsline}[3]{}
\bibliographystyle{apsrev4-2}
\bibliography{Refs_MainText}
\let\addcontentsline\oldaddcontentsline
\onecolumngrid
\newpage
\makeatletter
\renewcommand \thesection{S-\@arabic\c@section}
\renewcommand\thetable{S\@arabic\c@table}
\renewcommand \thefigure{S\@arabic\c@figure}
\renewcommand \theequation{S\@arabic\c@equation}
\makeatother
\setcounter{equation}{0}  
\setcounter{figure}{0}  
\setcounter{section}{0}  
{\begin{center}
	{\bf \large Supplemental Material for ``Point-Gap Bound States in Non-Hermitian Systems'' }
\end{center}}
\maketitle
\tableofcontents

\section{General theory of bound states in 1D non-Hermitian lattices}
\label{SecI}

In this section, we first present the general approach to bound states in the single-band non-Hermitian Hamiltonian, elaborating the derivation of the formulas Eq.(4) and Eq.(6) in the main text, and then extend these formulas to multi-band cases. 

\subsection{The bound states in single-band cases}
Consider a general one-band tight-binding model with the Hamiltonian in real space, 
\begin{equation}
	\begin{split}
		H_0 &= \sum_{r}\sum_{l=-m}^{n} t_{l} |r\rangle \langle r+l|  = 
		\sum_{k\in \mathrm{BZ}} \mathcal{H}(z\equiv e^{ik}) |k\rangle\langle k|,
	\end{split}
\end{equation}
where $m, n>0$ indicate the largest hopping range to right and left, respectively; $r$ represents the lattice site, and $t_{l}$ stands for the hopping parameter that only depends on the hopping range $l$ due to the translation symmetry. 
For the periodic boundary condition, $k \in \mathbb{R}$, and $z:=e^{ik}$ forms the unit circle $|z|=1$ in the complex $z$ plane. 
In general situations, $z$ is considered as complex variable and $\mathcal{H}(z)$ becomes Laurent polynomial of $z$, 
\begin{equation*}
	\mathcal{H}_0(z) =  \sum_{l=-m}^{n} t_l \, z^l  
	= \frac{t_n z^{m+n}+\dots+t_{-m}}{z^m} 
\end{equation*}
with $m$-order pole located at $z=0$. 

Starting from periodic boundary conditions (PBCs), we utilize Green's function method to derive the general form of bound states induced by single impurity, that is, Eq.(4) in the main text. 
Consider a single impurity at the center ($r=0$) of the periodic chain, 
\begin{equation}\label{Sec1_SingleImpurity}
	V =\lambda \, \sum\nolimits_r \delta(r)  |r\rangle \langle r|, 
\end{equation}
where $\lambda$ is the strength of impurity potential and assumes a complex value. 
The eigenequation for the total Hamiltonian reads, $(H_0+V)|\psi_{E}\rangle = E|\psi_{E}\rangle$, with $E$ the energy of eigenstate $|\psi_{E}\rangle$. 
The eigenstate whose energy does not belong to the PBC spectrum (denoted as $\sigma_{\mathrm{PBC}}$) can be further expressed as 
\begin{equation}\label{Sec1_EigenState}
	|\psi_E \rangle=\frac{1}{E-H_0}V|\psi_E \rangle=G_0(E)V|\psi_E \rangle,
\end{equation}
where $G_0(E)=(E-H_0)^{-1}$ represents the Green's function for $H_0$ under PBC, and $E\notin \sigma_{\mathrm{PBC}}$ avoids the singularity ($\det[E-H_0]\neq 0$). 
By substituting Eq.(\ref{Sec1_SingleImpurity}) into Eq.(\ref{Sec1_EigenState}) and taking spectral representation of the Green's function, we can obtain
\begin{equation}\label{Sec1_GenWavFun}
	\psi_E(r) = \langle r|\psi_{E}\rangle 
	=\frac{\lambda}{2\pi} \int_{-\pi}^{\pi} dk \frac{e^{i k r}}{E-\mathcal{H}_0(e^{ik})} \psi_{E}(0)
	=\lambda \psi_{E}(0) \oint_{\mathrm{BZ}} \frac{dz}{2\pi i z} \frac{z^r}{E-\mathcal{H}_0(z)}
	= \lambda \psi_{E}(0) \oint_{\mathrm{BZ}} \frac{dz}{2\pi i} \frac{z^{r+m-1}}{P_{E}(z)}.
\end{equation}
Here $P_{E}(z) = z^{m} (E-\mathcal{H}_0(z))$ is a non-negative-order polynomial of $z$, and the integral contour $\mathrm{BZ}$ is the unit circle$|z|=1$. 
Therefore, for given $E\notin \sigma_{\mathrm{PBC}}$, the contour integral in Eq.(\ref{Sec1_GenWavFun}) equals the sum of residues at the poles encircled by BZ ($|z|=1$). 
Note that these poles are exactly the zeros of $P_{\mathrm{E}}(z)$. 
Given that the magnitude of bound states will decay away from the impurity site ($r=0$), therefore, from Eq.(\ref{Sec1_GenWavFun}) the poles located in the interior (exterior) of the BZ are taken into account when $r>0$ ($r<0$). 
Finally, we can obtain the general form of bound states induced by single impurity under PBC, i.e., the Eq.(4) in the main text, 
\begin{equation}\label{Sec1_BoundStates}
	\psi_{\mathrm{BS}}(r) = \lambda \, \psi_{\mathrm{BS}}(0)
	\left\{  \begin{split}
		&\sum_{|z|<1} {\mathrm{R}}(E_{\mathrm{BS}}, z) \, z^{r}, \,\,\,\,\,\,\,\,\, r > 0; \\  
		&\sum_{|z|>1} - {\mathrm{R}}(E_{\mathrm{BS}}, z) \,  z^{r}, \,\,\,\, r < 0,
	\end{split}  \right.  
\end{equation}
where $R(E_{\mathrm{BS}},z)$ represents the residue of $[z(E_{\mathrm{BS}} - \mathcal{H}_0(z))]^{-1}$ at its pole $z$, and $E_{\mathrm{BS}}\notin \sigma_{PBC}$ indicates the bound-state energy. 
It is noteworthy that under PBC, impurity states with energies in the energy gaps ($E\notin \sigma_{PBC}$) have to be the bound states localized around the impurity position. 
In contrast, under open boundary condition (OBC), an impurity-induced energy level beyond the OBC continuum spectrum might be a localized edge state, which we will discuss later in Sec.~\ref{SecIV}. 
In addition, the relationship between impurity potential and bound-state energy, namely Eq.(6) in the main text, is derived from Eq.(\ref{Sec1_GenWavFun}) by adopting $r=0$, 
\begin{equation}\label{Sec1_LambdaSpec}
	\lambda^{-1}(E_\mathrm{BS}) = \oint_{\mathrm{BZ}} \frac{dz}{2\pi i} \frac{1}{z(E_{\mathrm{BS}} - \mathcal{H}_0(z))}= \sum\nolimits_{|z_i|<1} \mathrm{R}(E_\mathrm{BS}, z_i). 
\end{equation}

As boundary conditions transition from PBC to OBC, the spectral representation of Green function in Eq.(\ref{Sec1_GenWavFun}) will be changed. 
As a result, the integral contour is deformed from BZ to generalized Brillouin zone (GBZ). 
Correspondingly, Eq.(\ref{Sec1_GenWavFun}) becomes 
\begin{equation}\label{Sec1_GenBCs}
	\psi_E(r) = \lambda \psi_{E}(0) \oint_{\partial D} \frac{dz}{2\pi i} \frac{z^{r+m-1}}{P_{E}(z)},
\end{equation}
where $\partial D$ indicates the boundary of the open domain $D$ and can be interpreted as BZ, GBZ or intermediary trajectory between them, depending on the specific boundary conditions. 
Note that for given energy $E$, there are some poles $z$ in the complex plane. 
If the integral contour $\partial D$ is deformed but hasn't touched these poles yet, the integral (namely the wave function $\psi_E(r)$) is invariant. 
When the boundary conditions deviate from PBC, the Eq.(\ref{Sec1_BoundStates}) needs to be generalized as 
\begin{equation}\label{Sec1_GenBoundStates}
	\psi_{E}(r) = \lambda \, \psi_{E}(0)
	\left\{  \begin{split}
		&\sum_{z\in D} {\mathrm{R}}(E, z) \, z^{r}, \,\,\,\,\,\,\,\,\, r > 0; \\  
		&\sum_{z \notin D} - {\mathrm{R}}(E, z) \,  z^{r}, \,\,\,\, r < 0,
	\end{split}  \right.
\end{equation}
where $E$ does not belong to the continuum spectrum obtained from $\mathcal{H}_0(z \in \partial D)$. 
This formulation indicates that the localized behavior of bound states away from the impurity site is dominated by the poles with the largest magnitude within $D$ and the smallest magnitude outside $D$. 
Moreover, the relationship between impurity potential and impurity-state energy becomes 
\begin{equation}\label{Sec1_GenLambdaSpec}
	\lambda^{-1}(E) = \oint_{\mathrm{\partial D}} \frac{dz}{2\pi i} \frac{1}{z(E - \mathcal{H}_0(z))}= \sum\nolimits_{z\in D} \mathrm{R}(E, z). 
\end{equation}

\subsection{Extension to multi-band cases}
In this subsection, we extend Eq.(4) and Eq.(6) of the main text into multi-band cases. 
Generally, the free Hamiltonian and impurity potential can be generally expressed as 
\begin{equation}\label{Sec1_MultiBandHam}
	\begin{split}
		H_0 = \sum_{r,\alpha,\beta}\sum_{l=-m}^{n} t^{\alpha \beta}_{l} |r,\alpha\rangle \langle r+l,\beta|  
		= \sum_{k\in \mathrm{BZ}} \mathcal{H}_{\alpha\beta}(z) |k,\alpha\rangle\langle k,\beta| ;  \,\,\,\,\,\,
		V =\sum_{\alpha\beta} \lambda \mathcal{V}_{\alpha\beta}\, |0,\alpha\rangle \langle 0,\beta|,
	\end{split}
\end{equation}
where $|r,\alpha\rangle$ denotes the $\alpha$-th degree of freedom at the $r$-th unit cell. 
Suppose that the degree of freedom per unit cell is $d$, then $\mathcal{H}_0(z)$ is $d*d$ matrix. 
The characteristic polynomial is
\begin{equation*}
	f(E,z) = \det[E I_{d} - \mathcal{H}_0(z) ]
\end{equation*}
with $I_d$ a $d*d$ identity matrix. 
Therefore, the wave function with energy $E$ can be derived as 
\begin{equation*}
	\psi_{E,\alpha}(r) = \langle r, \alpha| \psi_E\rangle = \sum_{\beta,\sigma} \langle r,\alpha | G_0(E)|0,\beta\rangle \lambda \mathcal{V}_{\beta\sigma} \langle 0, \sigma|\psi_{E}\rangle 
	= \lambda \sum_{n=1}^d \sum_{\beta,\sigma} \oint_{\partial D} \frac{dz}{2\pi i} \frac{z^{r+m-1}}{E-E_n(z)} \mathcal{P}_{n,\alpha\beta}(z) \mathcal{V}_{\beta\sigma} \psi_{E,\sigma}(0),
\end{equation*}
where $\mathcal{P}_{n,\alpha\beta}(z)$ is defined as $\langle \alpha |u^{R}_n(z)\rangle \langle u^L_n(z)| \beta \rangle$, and $|u^{R/L}_n(z)\rangle$ represents the right/left eigenvector of matrix $E I_{d} - \mathcal{H}_0(z)$. 
Here the integral contour $\partial D$ depends on the boundary conditions, as mentioned before. 
Finally, one can obtain the form of impurity state in multi-band cases, 
\begin{equation}\label{Sec1_MBWavFunc}
	\psi_{E,\alpha}(r) = \lambda \, \sum_{\sigma} \psi_{E,\sigma}(0)
	\left\{  \begin{split}
		&\sum_{z\in D} {\mathrm{R}_{\alpha\sigma}}(E, z) \, z^{r}, \,\,\,\,\,\,\,\,\, r > 0; \\  
		&\sum_{z \notin D} - {\mathrm{R}_{\alpha\sigma}}(E, z) \,  z^{r}, \,\,\,\, r < 0
	\end{split}  \right.
\end{equation}
with $\mathrm{R}_{\alpha\sigma}(E,z)$ being the residue of function $\sum_{n,\beta} \mathcal{P}_{n,\alpha\beta}(z) \mathcal{V}_{\beta\sigma}/(E-E_{n}(z))$ at its poles. 
The form closely resembles Eq.(\ref{Sec1_GenBoundStates}) of the single-band case, and the similar results in multi-band case can be obtained. 

Generally, the Green's operator for the total Hamiltonian can be expanded as $G=G_0+G_0\Sigma G_0$, where $\Sigma = V+VG_0V+VG_0VG_0V+\dots$ and $V$ represents the impurity potential. 
Therefore, Eq.(6) in the main text can be extended into general multi-band cases by considering the pole of self-energy term~\cite{SM_GreenFuncBook}, 
\begin{equation*}
	\Sigma=V\sum_{n=0}^{\infty} (G_0V)^n  = V \frac{\mathbb{I}-(G_0V)^{\infty}}{\mathbb{I}-G_0 V} = \frac{V}{\mathbb{I}-G_0V}, \,\,\,\, 
	\mathrm{iff} \,\, |G_0V| < 1, 
\end{equation*}
where $|G_0V| < 1$ refers to the convergence condition. 
Finally, the relation between impurity potential and impurity-state energy can be obtained as
\begin{equation}
	\det[\mathbb{I} - G_0(E) V]=0,
\end{equation}
and $V$ has the form in Eq.(\ref{Sec1_MultiBandHam}) with the strength of impurity potential $\lambda$. 
So far, we have achieved the extension of Eq.(4) and Eq.(6) of the main text into multi-band cases. 

\section{Point gaps and bound states}
\label{SecII}

In this section, we first use an intuitive example to explain how to evaluate the minimum impurity potential $\Lambda_{\mathcal{E}_{i}}$ for the formation of bound states in the point gap $\mathcal{E}_i$ under PBC. 
Next, we provide more details on the two models presented in Fig.~2 of the main text and finally show an intricate example with a higher spectral winding number to validate the conclusions. 

\begin{figure}[t]
	\begin{center}
		\includegraphics[width=1\linewidth]{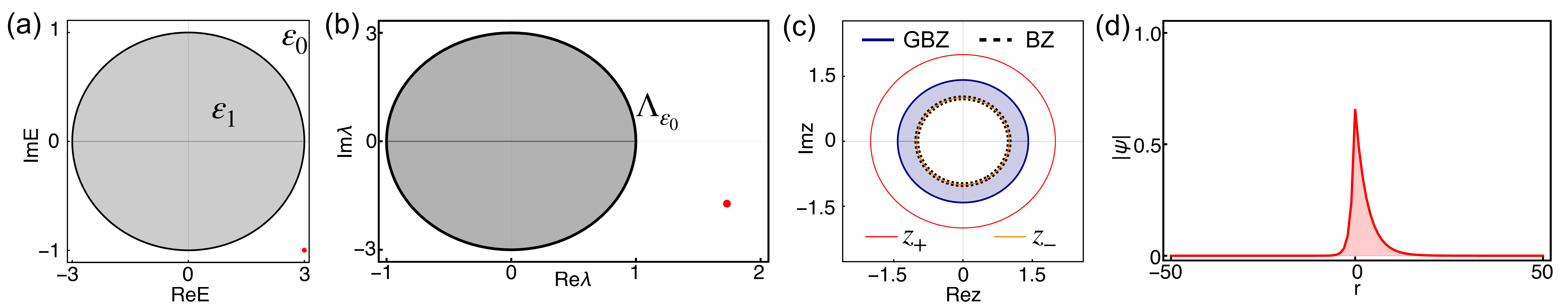}
		\par\end{center}
	\protect\caption{\label{Fig_SM1}
		The parameters in the Hatano-Nelson model are selected as $t_l=2$ and $t_r=1$. 
		(a) The PBC spectrum (the black curve), one bound-state energy $E=3.1$ (the red dot), and two point gaps, $\mathcal{E}_0$ with zero spectral winding number and $\mathcal{E}_1$ with $w_{\mathrm{BZ},\mathcal{E}_1}=1$. 
		(b) The minimum impurity potential $\Lambda_{\mathcal{E}_0}$ defined in Eq.(\ref{SM_HNLimitation}) is plotted by the black curve, and the impurity potential $\lambda=1.27$ to create the bound state in (a) is labeled by the red dot. 
		(c) The BZ (the black dashed circle), GBZ (the blue circle), and two zeros of $E-\mathcal{H}_0(z)$ with $E\in \sigma_{\mathrm{PBC}}$. 
		(d) The asymmetric spatial profile of the bound state.} 
\end{figure}

\subsection{The minimum impurity potentials}\label{sec:minimum threshold}
Here we use the Hatano-Nelson model, and the bulk Hamiltonian can be written in the non-Bloch from as
\begin{equation}\label{Sec2_HNModel}
	H_0(z) = t_r z^{-1} + t_l z,
\end{equation}
where $t_l\neq t_r$ indicates the hopping coefficient. 
As $z$ goes along BZ (the dashed circle in Fig.~\ref{Fig_SM1}(c)), $H_0(z)$ generates PBC spectrum $\sigma_{\mathrm{PBC}}$ (the black loop in Fig.~\ref{Fig_SM1}(a)). 
The PBC spectrum divides the complex energy plane into two disjoint regions, the point gap~\cite{SM_Kawabata2019PRX} $\mathcal{E}_0$ and $\mathcal{E}_1$ (the gray region). 
According to Eq.(\ref{Sec1_LambdaSpec}), the impurity potential required to produce the bound state with energy $E$ can be obtained. 
One conclusion in the main text states that the bound states cannot be created within point gaps that possess spectral winding of $n$ and $-m$, that is, $n=m=1$ in this example. 
Therefore, the bound-state energy cannot appear in the point gap $\mathcal{E}_1$ (the gray region in Fig.~\ref{Fig_SM1}(a)). 

\begin{figure}[b]
	\begin{center}
		\includegraphics[width=0.7\linewidth]{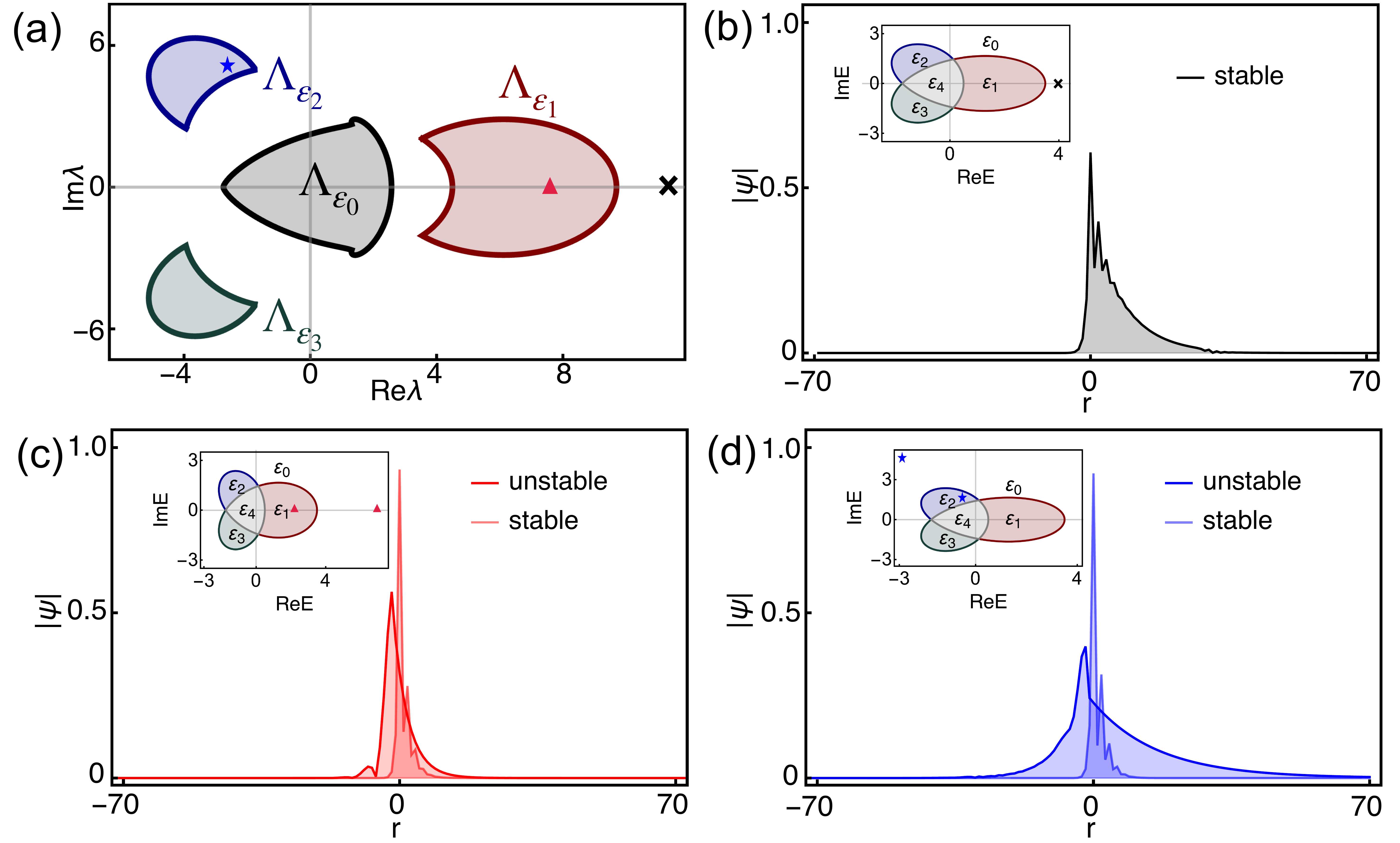}
		\par\end{center}
	\protect\caption{\label{Fig_SM2}
		The parameters in the Hamiltonian Eq.(\ref{Sec2_Model}) are chosen as $\left \{ t_{-2},t_{-1},t_1 \right\}=\left\{2, 1/2, 1 \right\}$. 
		(a) The $\lambda$ diagram for the Hamiltonian, where three sets of minimum impurity potentials are labeled by $\Lambda_{\mathcal{E}_{0,1,2,3}}$ and three representative $\lambda$ are selected as $3.298$ (the black cross), $6.948$ (the red triangle), and $-2.758+5.027i$ (the blue star). 
		For these three impurity potentials, the bound-state energies (with PBC spectrum) and spatial profile of bound states are shown in (b)-(d), respectively, where the stable (unstable) means that the bound state is insensitive (sensitive) to the boundary conditions. 
	} 
\end{figure}

As defined in the main text, the minimum impurity potentials relating to the point gap $\mathcal{E}_{i}$ has the form
\begin{equation}\label{Sec2_Limitation}
	\Lambda_{\mathcal{E}_i}:=\{ \lim\nolimits_{E_{\mathrm{BS}}\rightarrow E} \lambda(E_{\mathrm{BS}}) | E_{\mathrm{BS}} \in \mathcal{E}_i,  E \in \sigma_{\mathrm{PBC}} \},
\end{equation}
where $\lambda(E_{\mathrm{BS}})$ can be evaluated in Eq.(\ref{Sec1_LambdaSpec}). 
In this model, for each $E\notin \sigma_{\mathrm{PBC}}$, there are two zeros of $E-H_0(z)$, $z_\pm=(E \pm \sqrt{E^2-4t_l t_r})/2t_l$. 
The impurity potential can be further calculated as 
\begin{equation*}\label{Sec2_ModelImpPotential}
	\lambda(E_{\mathrm{BS}}) = -t_l (z_{-}-z_{+}) = \sqrt{E_{\mathrm{BS}}^2-4 t_l t_r}. 
\end{equation*}
Since the bound-state energy cannot appear in the point gap $\mathcal{E}_1$, only the minimum impurity potentials $\Lambda_{\mathcal{E}_{0}}$ for the point gap $\mathcal{E}_0$ can be obtained as $E_{\mathrm{BS}}\in \mathcal{E}_0$ approaches $\sigma_{\mathrm{PBC}}$. 
Finally, Eq.(\ref{Sec2_Limitation}) for this example can be calculated as 
\begin{equation}\label{SM_HNLimitation}
	\Lambda_{\mathcal{E}_0} =t_r e^{-i k}-t_l e^{i k }; \,\,\,\, k \in (0,2\pi],
\end{equation}
which is plotted by the black curve in Fig.~\ref{Fig_SM1}(b) for the parameters $t_l=2$ and $t_r=1$. 
When $\lambda$ is in the gray region surrounded by $\Lambda_{\mathcal{E}_0}$, it is insufficiently strong to give rise to bound states. 
As $\lambda$ surpasses the boundary $\Lambda_{\mathcal{E}_{0}}$, bound states first appear in the point gap $\mathcal{E}_0$. 
One bound state is shown in Fig.~\ref{Fig_SM1}(d). 
The corresponding bound-state energy is indicated by the red dot in (a) and the impurity potential $\lambda$ is labeled by the red dot in (b). 
Note that when $t_r=t_l =t \in \mathbb{R}$, the Hamiltonian reduces to Hermitian limit, and $\Lambda_{\mathcal{E}_0} = -2i t \sin{k} $ shrinks into a segment lying on the imaginary axis in the complex $\lambda$ plane. 
Therefore, one can always take an infinitesimal real impurity potential to confine bound states, as known in Hermitian systems. 

\begin{figure}[b]
	\begin{center}
		\includegraphics[width=0.7\linewidth]{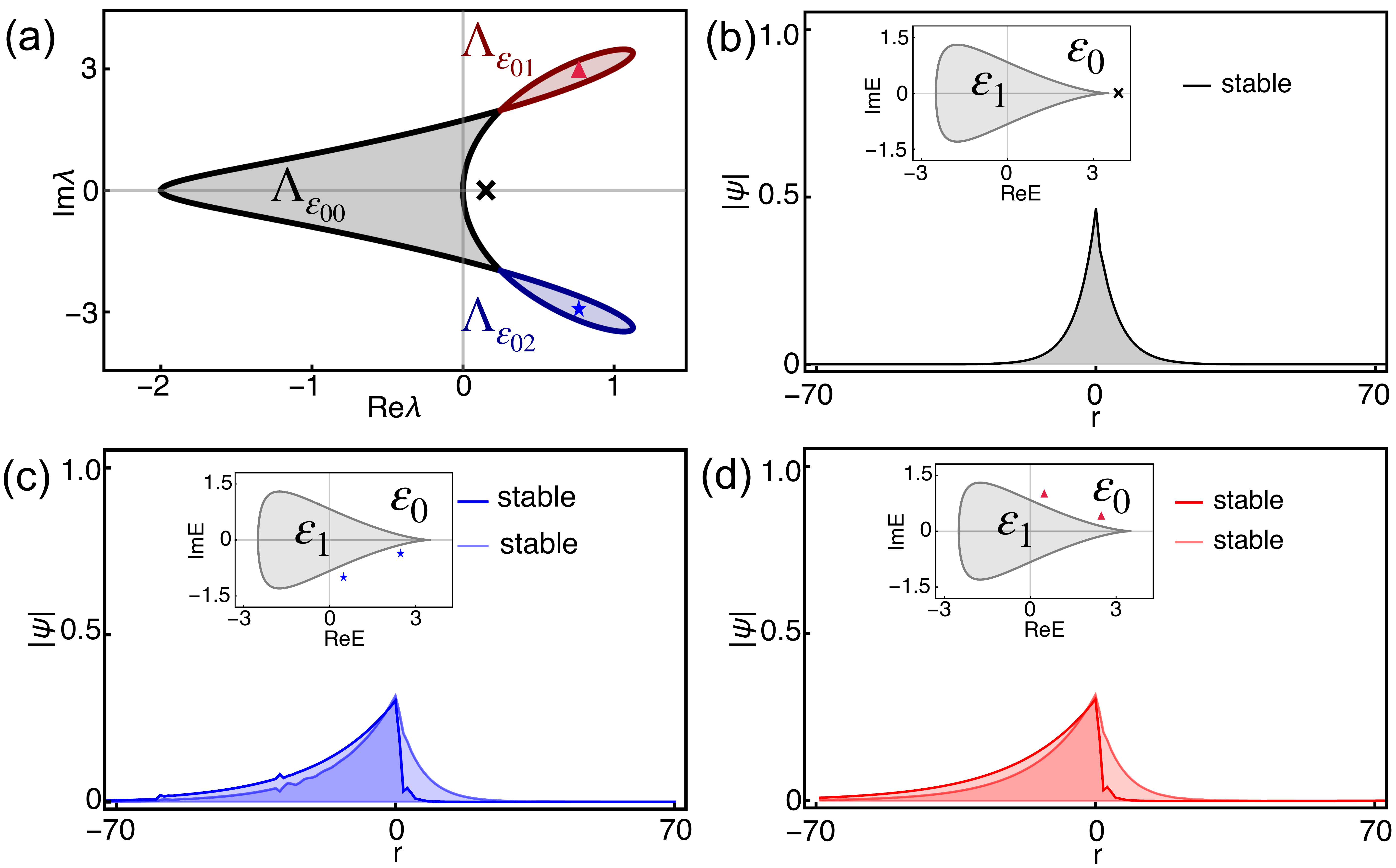}
		\par\end{center}
	\protect\caption{\label{Fig_SM3}
		The parameters in the model Eq.(\ref{Sec2_Model}) are set to be $\left \{ t_{-2},t_{-1},t_1 \right\}=\left\{1/2, 1, 2 \right\}$. 
		(a) shows the $\lambda$ diagram, where the minimum impurity potential for point gap $\mathcal{E}_0$, $\Lambda_{\mathcal{E}_0}$ defined in Eq.(\ref{SM_SecModelLimit}), are composed of three pieces labeled by $\Lambda_{\mathcal{E}_{00}}$, $\Lambda_{\mathcal{E}_{01}}$, and $\Lambda_{\mathcal{E}_{02}}$. 
		Three typical impurity potentials are chosen, that is, $\lambda=1$ (the black cross), $\lambda=0.8-2.9i$ (the blue star), and $\lambda=0.8+2.9i$ (the red triangle) shown in (a). 
		Correspondingly, the bound-state energies (with PBC spectrum) and spatial profile of bound states are shown in (b)-(d), respectively. }. 
\end{figure}

\subsection{The two examples in Fig.~2 of the main text}
In this section, we present more details about the examples used in Fig.~2 of the main text. 
The Hamiltonian can be expressed as
\begin{align}\label{Sec2_Model}
	\mathcal{H}_0(z)&= t_{-2}z^{-2} + t_{-1}z^{-1} +t_1 z.
\end{align}
The hopping parameters in the first example (Fig. 2(a)(c) in the main text) are chosen as $\left \{ t_{-2},t_{-1},t_1 \right\}=\left\{2, 1/2, 1 \right\}$. 
The $\lambda$ diagram can be evaluated by Eq.(\ref{Sec2_Limitation}) and plotted in Fig.~{\ref{Fig_SM2}}. 
We begin with the $\lambda$ diagram shown in Fig.~{\ref{Fig_SM2}}(a). 
There are several curves labeled by $\Lambda_{\mathcal{E}_i}$ corresponding to different point gaps $\mathcal{E}_i$ of the PBC spectrum.  
When $\lambda$ is in the gray region enclosed by $\Lambda_{\mathcal{E}_0}$, it is unable to create bound states. 
Therefore, we select three representative impurity potentials, marked by the black cross, the red triangle, and the blue star in Fig.~{\ref{Fig_SM2}}(a). 
The corresponding spatial profile of bound states and bound-state energies are shown in Fig.~\ref{Fig_SM2}(b)(c)(d) and the insets, respectively. 
When $\lambda$ is within the white region, only one bound state residing in the point gap $\mathcal{E}_0$ is generated, as shown in Fig.~{\ref{Fig_SM2}}(b). 
While $\lambda$ is chosen in the colored region, for example, red-colored (blue-colored) region, two bound states can be created, one in the point gap $\mathcal{E}_0$ and another in the $\mathcal{E}_1$ ($\mathcal{E}_2$), as shown in Fig.~{\ref{Fig_SM2}}(c)(d). 

The second example (Fig. 2(b)(d) in the main text) takes the parameters $\left \{ t_{-2},t_{-1},t_1 \right\}=\left\{1/2,1,2 \right\}$. 
The $\lambda$ diagram is shown in Fig.~\ref{Fig_SM3}(a), and the PBC spectrum of the example is plotted in the insets of Fig.~\ref{Fig_SM3}(b)-(d). 
Note that only the point gap $\mathcal{E}_0$ can accommodate bound-state energy, while $\mathcal{E}_1$ with the spectral winding number $w_{\mathrm{BZ},\mathcal{E}_1}=n=1$ cannot contain the bound-state energy. 
Therefore, the set of minimum impurity potentials in Eq.(\ref{Sec2_Limitation}) can only define for the point gap $\mathcal{E}_0$, 
that is, 
\begin{equation}\label{SM_SecModelLimit}
	\Lambda_{\mathcal{E}_0}:=\{ \lim\nolimits_{E_{\mathrm{BS}}\rightarrow E} \lambda(E_{\mathrm{BS}}) | E_{\mathrm{BS}} \in \mathcal{E}_0,  E \in \sigma_{\mathrm{PBC}} \},
\end{equation}
which is shown in Fig.~\ref{Fig_SM3}(a) by the curve composed of three curves connected by two self-intersections. 
We label the three components of $\Lambda_{\mathcal{E}_0}$ as $\Lambda_{\mathcal{E}_{00}}$, $\Lambda_{\mathcal{E}_{01}}$, and $\Lambda_{\mathcal{E}_{02}}$, respectively, and mark them by different colors. 
The $\lambda$ in the gray region contained by $\Lambda_{\mathcal{E}_{00}}$ is too weak to create bound states. 
We select three representative $\lambda$ marked by the black cross, the blue star, and the red triangle, as illustrated in Fig.~\ref{Fig_SM3}(a).
The corresponding bound-state energies and spatial profile are presented in Fig.~\ref{Fig_SM3}(b)-(d). 
There are two points different from the previous example. 
The first one is that when $\lambda$ is in the colored region (red or blue), two bound states appear in the point gap $\mathcal{E}_0$, and both are stable against the change of boundary conditions. 
Another point is that even though there exists the point gap $\mathcal{E}_1$ with nonzero spectral winding, an infinitesimal real impurity potential can still generate bound states, which is beyond the existing understanding and reveal the crucial role of Bloch saddle points (more details are discussed in Sec.~\ref{SecIII}). 

\begin{figure}[b]
	\begin{center}
		\includegraphics[width=0.7\linewidth]{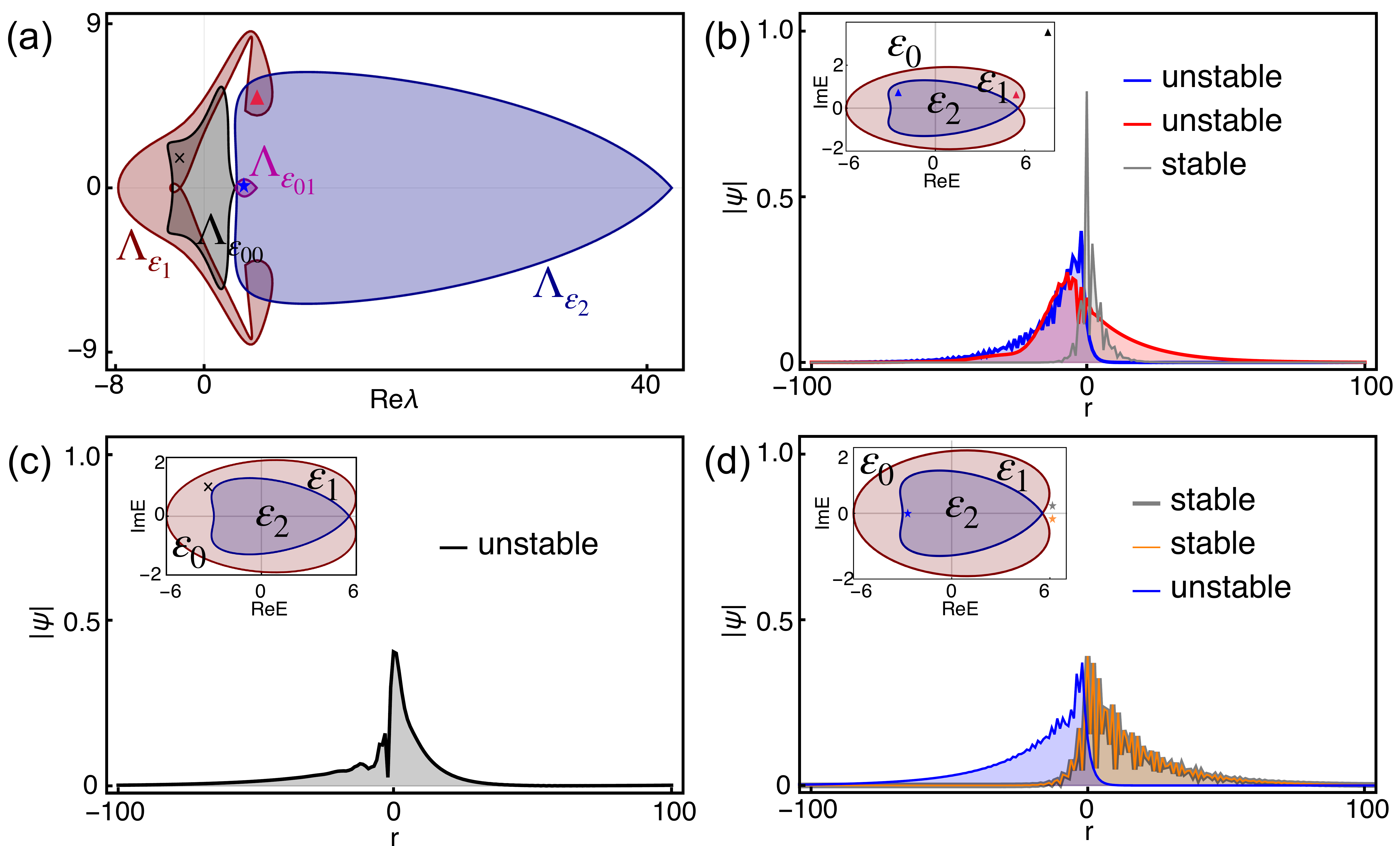}
		\par\end{center}
	\protect\caption{\label{Fig_SM4}
		(a) shows the $\lambda$ diagram for the Hamiltonian in Eq.(\ref{Sec3_Model}) and three typical impurity potentials, that is, $\lambda=5+5i$ (the red triangle), $\lambda=-2+2i$ (the black cross), and $\lambda=3.5$ (the blue star). 
		The corresponding bound-state energy (with PBC spectrum) and spatial profile of bound states are presented in (b)-(d), respectively. 
		Here, the bound states of energy in the point gap with nonzero (zero) spectral winding number are labeled as unstable (stable). 
	}
\end{figure}

\subsection{An example with higher spectral winding number}
Here an example with higher spectral winding number is presented, 
\begin{align}\label{Sec3_Model}
	\mathcal{H}_0(z) = t_{-3}z^{-3} + t_{-2}z^{-2} + t_{-1}z^{-1} + t_{1}z +t_{2}z^2 + t_{3}z^3,
\end{align}
and the parameters are chosen as $\left \{t_{-3},t_{-2},t_{-1},t_1,t_2,t_3\right\}=\left\{-3, -1/2, -1/2, -3/2, 3/2, 1 \right\}$. 
The PBC spectrum of the Hamiltonian is plotted in the inset of Fig.~\ref{Fig_SM4}(b)-(d). 
The complex energy plane is divided into three disjoint point gaps, labeled by $\mathcal{E}_{0,1,2}$, and the spectral winding number for these three point gaps can be calculated as $w_{\mathrm{BZ},\mathcal{E}_0}=0$, $w_{\mathrm{BZ},\mathcal{E}_1}=1$, and $w_{\mathrm{BZ},\mathcal{E}_2}=2$. 
Given the maximum hopping range to the left and right is $m=n=3$ in this example, all these three point gaps can potentially host bound-state energy according to our conclusion. 
The $\lambda$ diagram is shown in Fig.~{\ref{Fig_SM3}}(a). 
Based on Eq.(\ref{Sec2_Limitation}), there are three sets of minimum impurity potentials for the three point gaps, respectively, denoted as $\Lambda_{\mathcal{E}_{0}}$, $\Lambda_{\mathcal{E}_{1}}$, and $\Lambda_{\mathcal{E}_{2}}$. 
Note that $\Lambda_{\mathcal{E}_{0}}$ is composed of two pieces $\Lambda_{\mathcal{E}_{00}}$ and $\Lambda_{\mathcal{E}_{01}}$ linked by the self-intersection, similar to the case in Fig.~\ref{Fig_SM3}(a). 
Moreover, $\Lambda_{\mathcal{E}_{1}}$ and $\Lambda_{\mathcal{E}_{2}}$ can be obtained and correspond to point gaps $\mathcal{E}_{1}$ and $\mathcal{E}_{2}$, respectively. 
A noteworthy distinction from prior examples is the existence of overlapping regions. 
If $\lambda$ falls within these overlapping areas, bound states may appear in all associated point gaps. 

For example, when $\lambda=5+5i$ marked by the red triangle in Fig.~\ref{Fig_SM4}(a), there are three bound states appearing in the point gap $\mathcal{E}_0$, $\mathcal{E}_1$, and $\mathcal{E}_2$, respectively, as shown in Fig.~\ref{Fig_SM4}(b). 
Since the bound-state energy is forbidden in the gray region enclosed by $\Lambda_{\mathcal{E}_{00}}$ in Fig.~\ref{Fig_SM4}(a), the impurity with $\lambda=-2+2i$ (the black cross) only produces bound state in the point gap $\mathcal{E}_1$, as shown in Fig.~\ref{Fig_SM4}(c). 
When $\lambda=3.5$ is in the region included by $\Lambda_{\mathcal{E}_{01}}$, the impurity creates two bound states in the point gap $\mathcal{E}_0$, resembling to the case in Fig.~\ref{Fig_SM3}(a), and the third one bound state in the point gap $\mathcal{E}_2$ due to the overlap between the regions contained by $\Lambda_{\mathcal{E}_{01}}$ and $\Lambda_{\mathcal{E}_{2}}$, as shown in Fig.~\ref{Fig_SM4}(d). 
This intricate example demonstrates the function and importance of the $\lambda$ diagram. 
It's merely necessary to calculate the set of minimum impurity potentials $\Lambda_{\mathcal{E}_i}$ to fully know the number of bound states induced by the impurity potential $\lambda$ and the point gap where the bound-state energy resides. 

\section{Bloch saddle points and the threshold of impurity potential}
\label{SecIII}

The threshold of impurity potential refers to the least magnitude of impurity potential capable of generating bound states under PBC. 
Based on the formula in Eq.(\ref{Sec2_Limitation}) (namely, Eq.(8) in the main text), it can be inferred that the zero threshold emerges when $\Lambda_{\mathcal{E}_i}$ intersects at $\lambda=0$. 
In this section, we will prove that when the \emph{Bloch saddle point} (BSP) is present, a finite threshold of impurity potential is required to create the bound states; otherwise, an infinitesimal impurity potential can give rise to bound states (that is, the zero threshold of impurity potential). 

The saddle point $z_s$ refers to the point where the first derivative vanishes, $\partial_z \mathcal{H}_0(z)|_{z=z_s}=0$~\cite{SM_Longhi2019PRR,SM_ZSYangaGBZ,SM_YMHu2022}. 
Here, we define the BSP as the saddle point with a unit modulus specified as $|z_s|=1$. 
It's known that the OBC spectrum in 1D is always composed of some arcs in the complex energy plane~\cite{SM_Kai2020,SM_Okuma2020}.
At the endpoints of these arcs, at least two roots of $E-\mathcal{H}_0(z)=0$ merge. 
Consequently, the endpoints of the OBC spectrum always correspond to saddle points. 
However, the saddle point in the PBC spectrum is not a necessary ingredient. 

\begin{figure}[b]
	\begin{center}
		\includegraphics[width=0.9\linewidth]{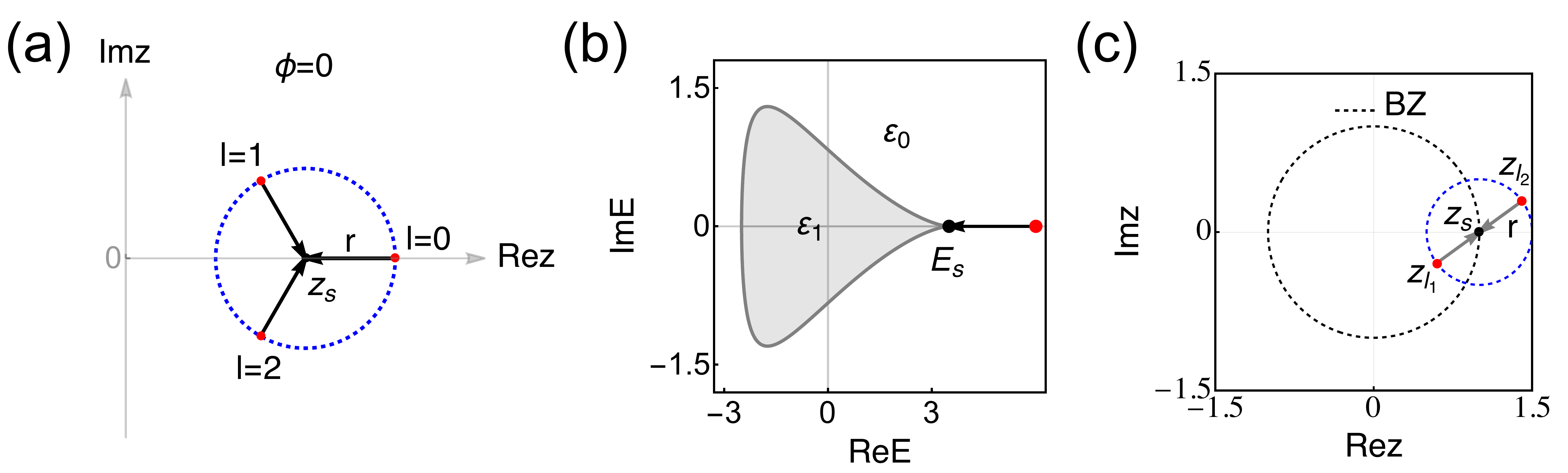}
		\par\end{center}
	\protect\caption{\label{Fig_SM5}
		(a) The illustration of three branches $z_l=z_s+r \exp(i(\phi+2\pi l)/n)$ with $n=3$ and $\phi=0$. 
		(b) The PBC spectrum (the gray curve) and BSP energy $E_s$ (the black dot) of model Hamiltonian in Eq.(\ref{Sec2_Model}). 
		(c) The BZ (the dashed unit circle) and two branches $z_{l_1}$ and $z_{l_2}$ approaching the saddle point $z_s$. }
\end{figure}

Since $\mathcal{H}_0(z)$ is analytic around the saddle point, the Hamiltonian can be expanded as, 
\begin{equation}
	\mathcal{H}_0(z)\approx \mathcal{H}_0(z_s)+\alpha(z-z_s)^n,\ n \geq 2\ \mathrm{and}\ n\in \mathbb{Z}
\end{equation}
where $\alpha \geq 0$ and $n$ denotes the order of the saddle point.
We can solve $z$ from the above equation,  
\begin{equation}
	z=z_s \left[ 1+\left(\frac{\mathcal{H}_0(z)-\mathcal{H}_0(z_s)}{\alpha z_s^n}\right)^{1/n} \right]
\end{equation} 
with $n$ branches around $z_s$ for a given energy $E=\mathcal{H}_0(z)$, that is, $z_l = z_s + r \exp(i (\phi+2\pi l)/n)$ with $l=0,1,\cdots, n-1$. 
We can approximately obtain the expression of $\mathcal{H}_0(z)$, 
\begin{equation}
	\mathcal{H}_0(z)=\mathcal{H}_0(z_s)+\alpha z_s^n r^n \exp(i\phi)
\end{equation}
with $r\geq 0$ and $0 < \phi \leq 2\pi$. 
Let $r\rightarrow 0$, $z_{l=0,\dots n-1}$ on $n$ different branches will finally collapse at the $n$-th order saddle point $z_s$. 
An example with $n=3$ is illustrated in Fig.~\ref{Fig_SM5}(a). 

Next, we will show that the presence of BSP leads to zero threshold of impurity potential. 
The Hamiltonian of the example is Eq.(\ref{Sec2_Model}) with parameters $\left \{ t_{-2},t_{-1},t_1 \right\}=\left\{1/2,1,2 \right\}$. 
As shown in Fig.~\ref{Fig_SM5}(b), the black dot $E_s=\mathcal{H}_0(z_s)$ denotes the BSP energy on the PBC spectrum (the gray curve). 
A neighboring energy (the red dot) can be expanded as $E=E_s+\alpha z_s^n r^n \exp(i\phi)$. 
Let $r \rightarrow 0$, $E$ is close to the BSP energy $E_s$. 
Correspondingly, two branches $z_{l_1}$ and $z_{l_2}$ separated on different sides of BZ will converge to the saddle point $z_s$, as shown in Fig.~\ref{Fig_SM5}(c). 
Therefore, only one branch $z_{l_1}$ will be counted into the integral Eq.(\ref{Sec1_GenWavFun}). 
Finally, from Eq.(\ref{Sec1_LambdaSpec}), one can deduce the impurity potential necessary to generate bound-state energies around BSP energy $E_s$, 
\begin{equation}\label{Sec3_SaddlePoints}
	\lambda^{-1}(E_s)=\lim_{r\rightarrow 0} \sum_{|z_i|<1}R(E,z_i),
\end{equation} 
where $R(E,z_i)\propto 1/\Pi_{i \neq j}(z_i-z_j)$. 
As $r\rightarrow 0$, $z_{l_1}, z_{l_2}$ approach the saddle point $z_s$, and the residue at $z_{l_1}$, that is, $R(E,z_{l_1}) \propto \lim_{r\rightarrow 0}1/(z_{l_1}-z_{l_2}) = \lim_{r\rightarrow 0} 1/ [r(e^{i \phi/2}-e^{i (\phi/2+\pi)})] \rightarrow \infty$, will dominate the right-hand side of Eq.(\ref{Sec3_SaddlePoints}). 
As a result, $\lambda(E_s)=0$ with $E_s=\mathcal{H}_0(z_s)$ and $z_s$ denotes BSP. 
For the general cases with $n>2$ branches, due to all branches distributed on a circle with the center at $z_s$ (for example, illustrated by Fig.~\ref{Fig_SM5}(a)), one can always find at least one branch inside the BZ and another outside, such that the residue at the inside branch will dominate the right-hand side of Eq.(\ref{Sec3_SaddlePoints}) and leads to $\lambda(E_s)=0$. 
So far, we have proved that the presence of BSP results in a zero threshold $\lambda(E_s)=0$, where $E_s$ represents the BSP energy. 

\section{The sensitivity of point-gap bound states to boundary conditions}\label{SecIV}

In this section, we first introduce the inverse of decay length for bound states, which enables us to understand the sensitivity of certain point-gap bound states to boundary conditions. 
Then we use an example in Eq.(\ref{Sec3_Model}) with a higher spectral winding number to demonstrate that the abrupt changes in decay length of the point-gap bound state correspond to the spectral winding number of the point gap, signifying bulk-edge correspondence in point-gap topology. 
Furthermore, as a supplement, we present how to obtain the unique intermediary trajectory $C_s$ between BZ and GBZ when the boundary condition is modulated by the boundary link strength $s$. 

\subsection{The inverse of decay length and the sensitivity of bound states to boundary conditions} 
\begin{figure}[t]
	\begin{center}
		\includegraphics[width=0.7\linewidth]{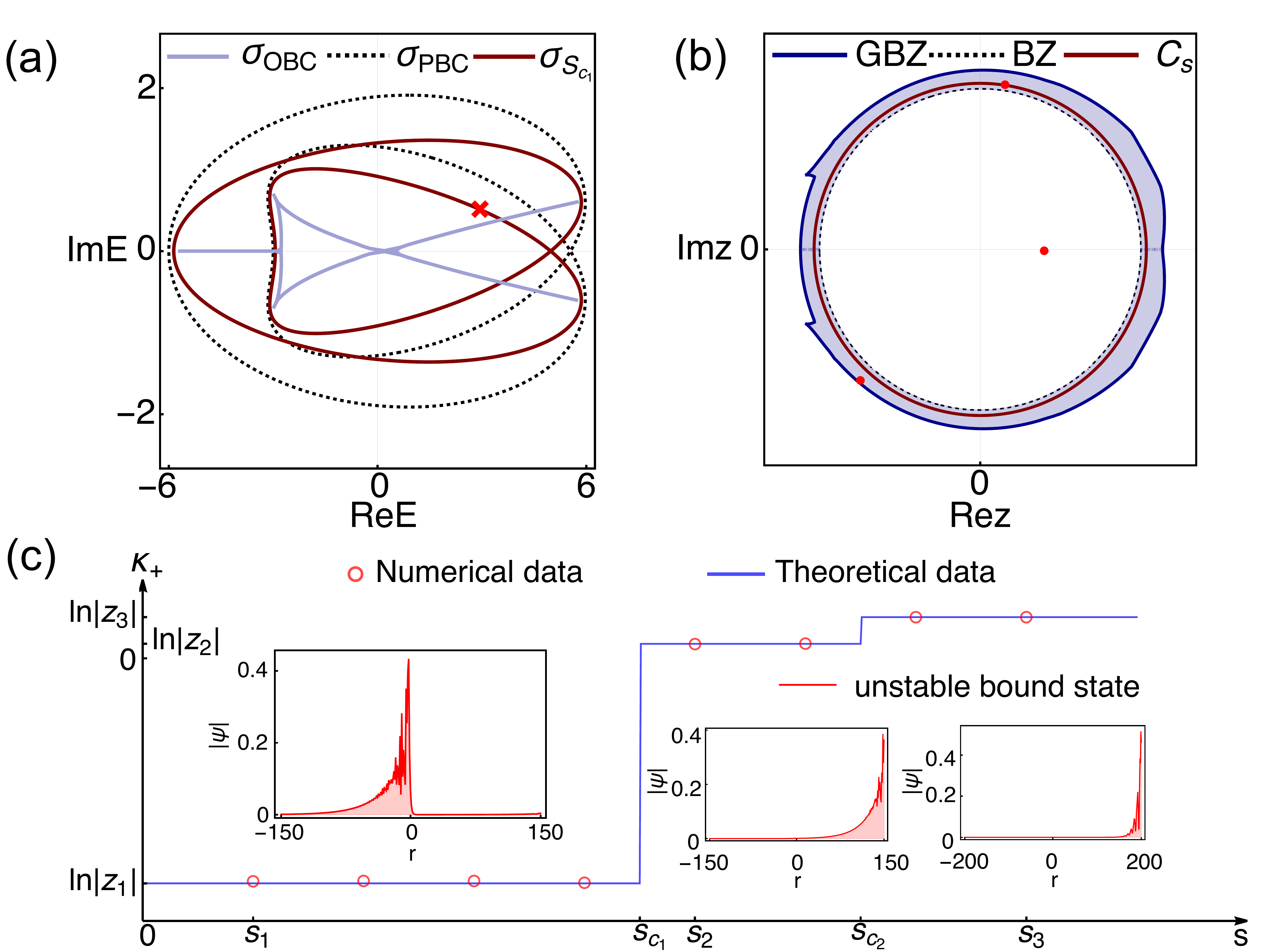}
		\par\end{center}
	\protect\caption{\label{Fig_SM6}
		(a) The spectra under different boundary conditions and the bound-state energy (the red cross). 
		(b) BZ, GBZ and critical trajectory $C_{s_{c_1}}$ that crosses one pole.
		The red dots indicate the poles of $\left[E_{\mathrm{BS}}-\right.$ $\left.\mathcal{H}_0(z)\right]^{-1}$ with $E_{\mathrm{BS}}$ represented by the red cross in (a), respectively. The three poles are ordered by their magnitude as $|z_1|<|z_2|<|z_3|$. 
		(c) The inverse of decay length $\kappa_+$ for the wave function with energy indicated by the red cross in (a) experiences two jumps. 
		The three insets show the spatial profile of bound state and edge modes with the boundary link strength $s_1=0.8$, $s_2=1-5^{-12}$ and $s_3=1-10^{-40}$, respectively. }
\end{figure}

We start with the tight-binding Hamiltonian under PBC and manipulate the boundary conditions via the boundary link strength $s$ ($0\leq s \leq 1$),
\begin{equation}\label{Sec4_BoundLink}
	H_0^s=H_0-s H_{B}
\end{equation}
where $H_0$ represents the Hamiltonian with PBC and $H_B$ denotes the boundary link term. 
As $s$ runs from $0$ to $1$, the spectrum of $H_0^s$ evolve from $\sigma_{\mathrm{PBC}}$ to $\sigma_{\mathrm{OBC}}$, and BZ is deformed into GBZ with the intermediary trajectory denoted as $C_s$. 
The corresponding intermediary spectrum can be obtained $\sigma_s:=\{\mathcal{H}_0(z)| z\in C_s, 0<s<1\}$. 
Generally, the Hamiltonian Eq.(\ref{Sec4_BoundLink}) can be written on the 1D ring geometry in real space as, 
\begin{equation}\label{Sec4_RealHam}
	H_0^s = \sum_{r=-N}^N \sum_{l=-m}^n t_l |r\rangle \langle r+l| - 
	s \left\{ \sum_{-m <d \leq 0} \sum_{l=-m}^{d-1} t_l |-N -d \rangle \langle -N-d + l | +  \sum_{ 0 \leq d <n }\sum_{l=d+1}^n t_l |N-d \rangle \langle N -d + l| \right\},
\end{equation}
where 1D ring geometry with $2N+1$ unit cell requires that $|-N - i \rangle \equiv | N-i+1 \rangle$. 
Using this notation, the first term in Eq.(\ref{Sec4_RealHam}) refers to the tight-binding Hamiltonian with system size $2N+1$ under PBC, and the second term represents the boundary link with the link strength $s$ in the range of $0\leq s \leq 1$. 
The example in Eq.(\ref{Sec3_Model}) is shown in Fig.~\ref{Fig_SM6}, the largest hopping range of this example is $m=n=3$, and the system parameters are chosen as $\left \{t_{-3},t_{-2},t_{-1},t_0,t_1,t_2,t_3\right\}=\left\{-3, -0.5, -0.5, 0, -1.5, 1.5, 1 \right\}$. 
The PBC ($\sigma_{\mathrm{PBC}}$), OBC ($\sigma_{\mathrm{OBC}}$), and an intermediary spectrum ($\sigma_{s}$, $s=s_{c_1}$) are shown by the black unit circle, the blue and red curve in Fig.~\ref{Fig_SM6}(a), respectively. 

After introducing the impurity potential, the total Hamiltonian reads 
\begin{equation}\label{Sec4_TotalHam}
	H^s = H_0^s + \lambda |0\rangle\langle 0|.
\end{equation}
As we discussed previously, the impurity will induce bound states under PBCs, and the form of bound state is presented in Eq.(\ref{Sec1_GenBoundStates}), where the integral contour $\partial D$ will be replaced by the trajectory $C_s$ in this scenario. 
When the bound-state energy $E_{\mathrm{BS}}$ resides in the point gap with nonzero spectral winding number, correspondingly, the zeros of $E_{\mathrm{BS}}-\mathcal{H}_0(z)$ distribute in the intermediary region between BZ and GBZ. 
For example, $E_{\mathrm{BS}}=3+0.5i$ marked by the red cross in Fig.~\ref{Fig_SM6}(a) stays in the point gap $\mathcal{E}_2$ with spectral winding number $w_{\mathrm{BZ},\mathcal{E}_2}=2$. 
Correspondingly, there are two zeros (the red dots) in the intermediary region between BZ and GBZ (the blue shaded region), as shown in Fig.~\ref{Fig_SM6}(b). 
As boundary condition transitions from PBC to OBC, the contour $C_s$ passes through these two zeros, the decay length of corresponding wave function from Eq.(\ref{Sec1_GenBoundStates}) will experience two abrupt jumps (Fig.~\ref{Fig_SM6}(c)) since the number of poles contained by $C_s$ is changed. 

The inverse of decay length for the impurity state on the right and left sides of the impurity can be defined as
\begin{equation}
	\begin{split}
		&\kappa_+ := \partial_r \ln|\psi_{E}(r)| \,\,\,\, r>0; \\
		&\kappa_- := \partial_r \ln|\psi_{E}(r)|  \,\,\,\,  r<0.
	\end{split}
\end{equation}
It can be derived from Eq.(\ref{Sec1_GenBoundStates}) that the localization behavior of the wave function in the $r>0$ ($r<0$) region is dominated by the pole $\tilde{z}_{\mathrm{in}}$ ($\tilde{z}_{\mathrm{out}}$) inside (outside) $C_s$, where $|\tilde{z}_{\mathrm{in}}|$ ($|\tilde{z}_{\mathrm{out}}|$) represents the largest (smallest) magnitude of the poles within (out) the integral contour $C_s$.  
Therefore, it follows that 
\begin{equation}
	\kappa_+ \approx \ln |\tilde{z}_{\mathrm{in}}|; \,\,\,\, 
	\kappa_- \approx \ln |\tilde{z}_{\mathrm{out}}|.
\end{equation}
In Fig.~\ref{Fig_SM6}(c), we choose eight representative boundary link strengths $s$ and calculate the corresponding $\kappa_+$ for the wave function with energy $E=3+0.5i$ (the red cross in Fig.~\ref{Fig_SM6}(a)). 
The value of $\kappa_+$  for different boundary link $s$ are denoted by the red hollow circles in Fig.~\ref{Fig_SM6}(a), which agrees well with the dominant poles (the three blue platforms) enclosed by $C_s$. 
Here $z_1$, $z_2$, $z_3$ are the three poles denoted by the red dots in Fig.~\ref{Fig_SM6}(b) and ordered by their magnitude $|z_1|<|z_2|<|z_3|$. 
During the transition of boundary conditions, $C_s$ passes through two poles, correspondingly, $\kappa_+$ experiences two jumps ($s=s_{c_1}$ and $s_{c_2}$), as shown in Fig.~\ref{Fig_SM6}(c). 
The spatial profile of impurity states at $s=s_1,s_2,s_3$ with the energy $E=3+0.5i$ (marked by the red cross in Fig.~\ref{Fig_SM6}(a)) are presented in the insets of Fig.~\ref{Fig_SM6}(c). 
One remarkable point in Fig.~\ref{Fig_SM6}(c) is that after the first jump, $k_+$ is greater than zero, which means the wave function is no longer a bound state around the impurity site while becomes localized edge mode, exhibiting sensitivity (or instability) to change of boundary conditions. 

\subsection{The intermediary trajectory $C_s$ between BZ and GBZ}
Here, we will discuss how to determine the intermediary trajectory $C_{s}$ and demonstrate the uniqueness of $C_{s}$ for given boundary link strength $s$ in the Hamiltonian Eq.(\ref{Sec4_RealHam}). 

As the boundary link $s$ is modulated from $0$ to $1$, the PBC spectrum is continuously deformed into OBC spectrum. 
Therefore, the intermediary spectrum $\sigma_s$ is always closed curve in the complex energy plane. 
Since $\mathcal{H}_0(z)$ is analytic in the region between the BZ and GBZ, one can always find a closed curve $C_s$ between the BZ and GBZ such that $\sigma_s$ is the image of $C_s$ under $\mathcal{H}_0(z)$. 

\begin{figure}[t]
	\begin{center}
		\includegraphics[width=0.7\linewidth]{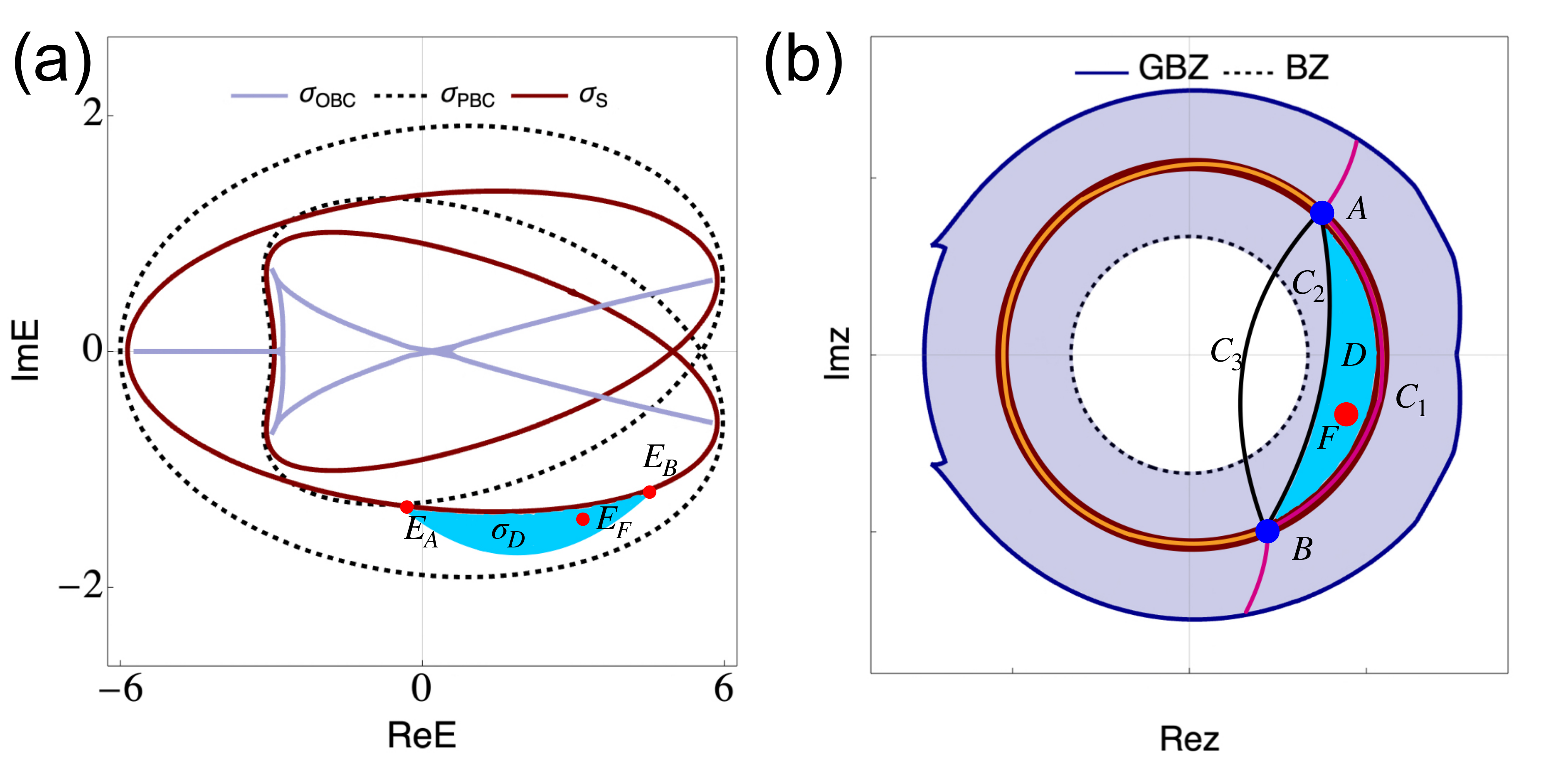}
		\par\end{center}
	\protect\caption{\label{Fig_SM7}
		(a) The spectra under different boundary conditions. 
		(b) BZ, GBZ, and three types of closed trajectories that include path $C_1$, $C_2$, and $C_3$, respectively. 
		The blue points A, B and the red point F in (b) are the solutions of the function $E-\mathcal{H}_0(z)=0$ with energies taking $E_A$, $E_B$ and $E_F$ in (a). 
		The blue region D in (b) is mapped under $\mathcal{H}_0(z)$ into the blue region $\sigma_D$ in (a). }
\end{figure}

When $E$ goes along the intermediary spectrum $\sigma_s$ (the red curve in Fig.~\ref{Fig_SM7}(a)), the solutions of $E-\mathcal{H}_0(z)=0$ form a couple of closed curves in the complex $z$ plane. 
Among these closed curves, only one is selected to be $C_s$. 
There are two criteria for the intermediary trajectory $C_s$: 
(i) $C_s$ is a continuous and closed curve in the region between BZ and GBZ; 
(ii) the region between BZ and $C_s$ exactly corresponds to the area swept by $\sigma_s$ from PBC spectrum $\sigma_{\mathrm{PBC}}$ in the complex energy plane. 
The criterion (i) comes from the fact that the intermediary spectrum $\sigma_s:=\{\mathcal{H}_0(z)| z\in C_s\}$ is a continuous and closed curve in the complex energy plane. 
Using these two conditions, one can determine the unique trajectory $C_{s}$ for given boundary link $s$. 
One illustration is shown in Fig.~\ref{Fig_SM7}(b). 
The pre-images of $\sigma_s$ under $\mathcal{H}_0(z)$ form some continuous and closed trajectories, which are ramified into three paths $C_1$, $C_2$, and $C_3$ between the branch points $A$ and $B$. 
The path $C_3$ does not satisfy criterion (i) and can be excluded first. 
The closed trajectories including paths $C_1$ and $C_2$ encompass differing areas, with the discrepancy represented by the blue region in Fig.~\ref{Fig_SM7}(b). 
The point $F$ in the blue region can be mapped into $E_F$ in the complex energy plane, which has been swept by $\sigma_s$. 
Therefore, according to criterion (ii), the path $C_1$ should be included in the trajectory $C_s$. 
Finally, we can determine the unique trajectory $C_s$ corresponding to spectrum $\sigma_s$ in Fig.~\ref{Fig_SM7}(a) to be the orange curve in Fig.~\ref{Fig_SM7}(b).

\end{document}